\documentclass[11pt, reqno]{amsart}
\usepackage{cite}
\usepackage{amssymb}   
\usepackage{amsfonts}   
\usepackage{amsthm, graphicx, color}  
\usepackage{tikz}
\usetikzlibrary{shapes.geometric, arrows}             
\usepackage[colorlinks=true, linkcolor=blue, citecolor=blue, urlcolor=blue]{hyperref}
\label{key}
\usepackage{framed}
\usepackage{mathrsfs}
\usepackage[english]{babel}
\usepackage{orcidlink} 
\allowdisplaybreaks
\newtheorem{thm}{Theorem}[section]

\theoremstyle{definition}
\newtheorem{definition}[thm]{Definition}
\theoremstyle{remark}
\newtheorem{remark}[thm]{Remark}
\numberwithin{equation}{section}

\DeclareMathSymbol{\subsetneqq}{\mathbin}{AMSb}{36}

\begin{document}
	
	\begin{flushleft}
		{\bf\Large {A Novel Two-Dimensional Wigner Distribution Framework via the Quadratic Phase Fourier Transform with a Non-Separable Kernel}}
	\end{flushleft}
	
	\parindent=0mm \vspace{.4in}

{\bf{Mukul Chauhan$^{ \orcidlink{0009-0009-1332-3005}}$, Waseem Z. Lone$^{ \orcidlink{0000-0002-9826-9475}}$, Amit K. Verma$^{*,\orcidlink{0000-0001-8768-094X}}$}}

\parindent=0mm \vspace{.1in}
{\small \it Department of Mathematics, Indian Institute of Technology Patna, Bihta, Patna 801103, (BR) India. \\		
	
	E-mail: $\text{mukul\_2421ma13@iitp.ac.in}$;\,$\text{lone.waseem.393@gmail.com}$;\,$\text{akverma@iitp.ac.in}$ }\\

$^{\star}$Corresponding author: $\text{akverma@iitp.ac.in}$ 

\parindent=0mm \vspace{.2in}
{\small {\bf Abstract.} This paper introduces a novel time-frequency distribution, referred to as the Two-Dimensional Non-Separable Quadratic Phase Wigner Distribution (2D-NSQPWD), formulated within the framework of the Two-Dimensional Non-Separable Quadratic Phase Fourier Transform (2D-NSQPFT). By replacing the classical Fourier kernel with the NSQPFT kernel, the proposed distribution generalizes the classical Wigner distribution and effectively captures complex, non-separable signal structures. We rigorously establish several key properties of the 2D-NSQPWD, including time and frequency shift invariance, marginal behavior, conjugate symmetry, convolution relations, and Moyal's identity. Furthermore, the connection between the 2D-NSQPWD and the two-dimensional short-time Fourier transform (2D-STFT) is explored. The distribution's effectiveness is demonstrated through its application to single-, bi-, and tri-component two-dimensional linear frequency modulated (2D-LFM) signals, where it shows superior performance in cross-term suppression and signal localization.
	
	\parindent=0mm \vspace{.1in}
	
	\parindent=0mm \vspace{.1in}
	}

\section{Introduction}\label{S1} 
The quadratic-phase Fourier transform (QPFT), proposed by Castro and collaborators\cite{Castro1,Castro2}, offers a unified framework that encompasses several classical transforms such as the Fourier transform (FT), fractional Fourier transform (FrFT), and linear canonical transform (LCT) through its generalized quadratic phase structure. Its ability to accommodate both dynamic and stable signal components makes the QPFT a powerful and versatile tool in modern signal analysis. Moreover, its inherent flexibility in handling multiple adjustable parameters enables its application across a wide range of fields, including harmonic analysis, sampling theory, image and signal processing, and various other domains in science and engineering \cite{Prasad,Shah1,Shah2,AK}.

\parindent=8mm\vspace{.1in}	
Recently, Castro et al. \cite{CMA} introduced a multidimensional extension of the QPFT, significantly advancing its analytical capabilities and expanding its applicability to a broader range of multidimensional problems. By replacing traditional scalar parameters with real valued matrices, the proposed framework allows for greater adaptability to complex signal structures while preserving essential properties such as invertibility and energy conservation. In the two-dimensional case, this formulation yields fifteen independent parameters, offering enhanced flexibility compared to classical transforms. Importantly, the multidimensional QPFT unifies several well known operations including 2D-FT, 2D-FrFT, 2D-LCT, affine transformations under a single, generalized structure, making it a powerful and versatile tool for advanced time–frequency analysis in theoretical and applied contexts\cite{Bahri1,Zayed1,Pei1,Dong,Zhao,Mukul}.

\parindent=8mm\vspace{.1in}	
In the two-dimensional setting, the NSQPFT is governed by a matrix tuple $\Omega$ composed of five real $2 \times 2$ matrices, defined as
\begin{align}\label{S1E1}
	\Omega = \Bigg\{ & A = 
	\begin{bmatrix}
		a_{11} & a_{12} \\
		a_{21} & a_{22}
	\end{bmatrix}, \,
	B = 
	\begin{bmatrix}
		b_{11} & b_{12} \\
		b_{12} & b_{22}
	\end{bmatrix}, \, \det(B) \ne 0, \,
	C = 
	\begin{bmatrix}
		c_{11} & c_{12} \\
		c_{21} & c_{22}
	\end{bmatrix}, \,
	D = 
	\begin{bmatrix}
		d_{11} & d_{12} \\
		d_{21} & d_{22}
	\end{bmatrix}, \notag\\
	& ~~E = 
	\begin{bmatrix}
		e_{11} & e_{12} \\
		e_{21} & e_{22}
	\end{bmatrix}
	\Bigg\}.
\end{align}

Assuming $\Omega$ is given by \eqref{S1E1}, the 2D-NSQPFT of a function $f \in L^2(\mathbb{R}^2)$ is expressed as
\begin{align}\label{S1E2}
	\mathcal{Q}_{\Omega} \big[f\big](\boldsymbol{\omega})& = \int_{\mathbb{R}^2} f({\bf x})\, \mathcal{K}_{\Omega}({\bf x,\boldsymbol{\omega}})\,d{\bf x},
\end{align}
where the transformation kernel is explicitly given by
\begin{align}\label{S1E3}
	\mathcal{K}_{\Omega}({\bf x,\boldsymbol{\omega}}) &= \frac{i\,\sqrt{\det(B)}}{2\pi}\,\exp\Big\{i\big(\boldsymbol{\omega}^{T} A \boldsymbol{\omega} +\boldsymbol{\omega}^{T} B {\bf x} +  {\bf x}^T C {\bf x} +  \vec{1} D \boldsymbol{\omega}+ \vec{1} E {\bf x}\big) \Big\} 
\end{align}
with $B$ being a symmetric matrix,  $\vec{1}= (1, 1)^T $, $\boldsymbol{\omega}=(\omega_1,\omega_2)^T $ and ${\bf x}=(x_1,x_2)^T$. Moreover, for any square-integrable function $	f({\bf{x}})$, the inverse of the MDQPFT is given by
\begin{align}
	f({\bf{x}}) =	\mathcal{Q}_{\Omega}^{-1}\Big(\mathcal{Q}_{\Omega} \big[f\big](\boldsymbol{\omega})\Big)({\bf{x}})= \overline{\Lambda(B, n)} 
	\int_{\mathbb{R}^n} 
	\mathcal{Q}_{\Omega}[f](\boldsymbol{\omega})\, 
	\overline{\mathcal{K}_{\Omega}(\boldsymbol{\omega},{\bf x})} \, 
	d\boldsymbol{\omega}.
\end{align}

\parindent=8mm\vspace{.1in}	
The Wigner distribution (WD), originally introduced by Wigner in $1932$ for quantum mechanics, was later adapted by Ville in $1948$ for signal processing applications \cite{Wigner, Ville}. In one dimension, it gained prominence due to its excellent time–frequency resolution, making it particularly effective for analyzing nonstationary signals. However, its practical utility was hindered by the presence of cross-term interference in multicomponent signals. To overcome this limitation, more advanced frameworks such as Cohen’s class were developed, introducing smoothing kernels that effectively reduce cross-terms while preserving the essential features of the original signal \cite{Cohen,John}.

\parindent=8mm\vspace{.1in}
The extension of the WD to two dimensions uncovers significant structural distinctions. For separable signals, it elegantly reduces to a product of 1D distributions, ensuring energy localization and interpretability. However, for non-separable (NS) signals such as those exhibiting affine or rotational features cross-terms become dominant, complicating analysis. To address this, the affine Wigner distribution, introduced by Bertrand~\cite{Bertrand}, aligns with the geometric characteristics of such signals, improving clarity and representation. Among the classical tools in 2D-signal analysis, the 2D-WD stands out for its effectiveness, particularly in detecting and estimating 2D-linear frequency-modulated (2D-LFM) components. As formulated by Pei and Ding~\cite{Pei2}, the 2D-WD of a signal $f(x, y)$ is defined as:

\parindent=0mm\vspace{.1in}
\begin{align}\label{S1E4}
	\mathcal{W}_f(x_1, x_2, \omega_1, \omega_2) &= \int_{\mathbb{R}^2} f\left(x_1 + \frac{\tau}{2}, x_2 + \frac{\eta}{2}\right)
	\overline{f\left(x_1 - \frac{\tau}{2}, x_2 - \frac{\eta}{2}\right)} e^{-i\,(\omega_1\tau + \omega_2\eta)} \, d\tau \, d\eta.
\end{align}

\parindent=8mm\vspace{.1in}
While the 2D-WD has shown remarkable effectiveness in traditional signal analysis frameworks. In recent years, its integration with advanced transforms such as the FrFT and the LCT has attracted growing interest particularly for applications involving non-separable signal structures. Numerous studies have highlighted the advantages of merging the 2D-WD with these generalized transforms, citing their increased flexibility in representing and analyzing intricate signal behaviors \cite{Sahin,Zhang,Teali }. This synergy has led to improved adaptability and better localization in time-frequency representations.

\parindent=8mm\vspace{.1in}
However, a critical gap persists in extending the 2D-WD into the domain of the 2D-NSQPFT a framework that provides even greater generalization and adaptability. As far as current literature indicates, no formal attempt has been made to formulate the 2D-WD within the 2D-NSQPFT context. Motivated by the successes of earlier generalizations with FrFT\cite{Zayed2} and LCT\cite{Minh1}, this paper introduces a novel formulation of the 2D-WD by substituting the conventional Fourier kernel with the kernel of the 2D-NSQPFT:
\begin{align}\label{S1E5}
	\mathcal{W}_f(x_1, x_2, \omega_1, \omega_2) &= \int_{\mathbb{R}^2} f\left(x_1 + \frac{\tau}{2}, x_2 + \frac{\eta}{2}\right)
	\overline{f\left(x_1 - \frac{\tau}{2}, x_2 - \frac{\eta}{2}\right)}\, \mathcal{K}_{\Omega}(\tau,\eta,\omega_1, \omega_2) \, d\tau \, d\eta.
\end{align}

\parindent=8mm\vspace{.1in}
The proposed distribution preserves the fine resolution of the classical 2D-Wigner distribution while integrating the structural adaptability of the 2D-NSQPFT, enabling effective analysis of non-separable and geometrically complex signals. The main contributions of this work are threefold: (i) the introduction of the 2D-NSQPWD, a unified and generalized time–frequency representation that includes both the classical 2D-Wigner distribution and the gyrator-Wigner distribution as special cases; (ii) a rigorous investigation of the fundamental theoretical properties of the proposed distribution; and (iii) extensive graphical simulations demonstrating the efficacy of the 2D-NSQPWD in analyzing single-, bi-component, and tri-component 2D-LFM signals.

\parindent=8mm\vspace{.1in}
The remainder of this article is organized as follows. Section \ref{S2} provides a brief overview of the foundational concepts, including the 2D-NSQPFT  and the 2D-WD. Section \ref{S3} presents the formulation of the novel 2D-WD based on the NSQPFT framework, along with a thorough investigation of its core theoretical properties. In Section \ref{S4}, we explore the practical effectiveness of the proposed distribution in analyzing single, bi, and tri component 2D-LFM signals, with particular emphasis on its superior cross-term suppression capability. Finally, Section \ref{S5} concludes the paper by summarizing the key contributions and insights derived from this work.

   \section{2D-WD in the quadratic phase Domain }\label{S2}

\parindent=0mm
To construct the 2D-NSQPFT and its corresponding kernel function, we first outline the fundamental notations, define the relevant parameters, and establish the underlying mathematical framework. The kernel formulation in \eqref{S1E3}, when expanded component-wise using the parameters specified in \eqref{S1E1}, yields the following expression:
\begin{align*}
    &\mathcal{K}_{\Omega}(x_1, x_2,\omega_1, \omega_2)\notag\\
	&= \frac{i\,\sqrt{\det(B)}}{2\pi} \, \exp \Big\{ i \big(a_{11} \omega_1^2 + (a_{12}+a_{21}) \omega_1 \omega_2 + a_{22} \omega_2^2+   (d_{11}+d_{21}) \omega_1 + (d_{12}+d_{22}) \omega_2 \Big) \Big\}\\
	&\qquad\times	\exp \Big\{ i \big( c_{11} x_1^2 + (c_{12}+c_{21}) x_1 x_2 + c_{22} x_2^2 + (e_{11}+e_{21}) x_1 + (e_{12}+e_{22}) x_2 
	\big) \Big\}\\
	&\qquad\times\exp \Big\{ i\big( x_1	(\omega_1b_{11} +\omega_2b_{12})  + x_2(b_{12} \omega_1 + b_{22}  \omega_2)\big)\Big\}.
\end{align*}

Define the constants as	
\begin{align}\label{S2E1}
	\begin{bmatrix}
		k_1 \\ k_2 \\ k_3 \\ k_4 \\ k_5
	\end{bmatrix}
	=
	\begin{bmatrix}
		a_{11} \\
		a_{12} + a_{21} \\
		a_{22} \\
		d_{11} + d_{21} \\
		d_{12} + d_{22}
	\end{bmatrix},
	\quad
	\begin{bmatrix}
		m_1 \\ m_2 \\ m_3 \\ m_4 \\ m_5
	\end{bmatrix}
	=
	\begin{bmatrix}
		c_{11} \\
		c_{12} + c_{21} \\
		c_{22} \\
		e_{11} + e_{21} \\
		e_{12} + e_{22}
	\end{bmatrix}.
\end{align}
Then, the kernel $\mathcal{K}_{\Omega}(x_1, x_2,\omega_1, \omega_2)$ given by \eqref{S1E3} becomes
\begin{align*}
	\mathcal{K}_{\Omega}(x_1, x_2,\omega_1, \omega_2)
	&= \frac{i\,\sqrt{\det(B)}}{2\pi}\exp \Big\{ i \big( 
	k_1 \omega_1^2 + k_2 \omega_1 \omega_2 + k_3 \omega_2^2 
	+ k_4 \omega_1 + k_5 \omega_2 
	\big) \Big\} \notag\\
	&\quad \times \exp \Big\{ i \big( 
	m_1 x_1^2 + m_2 x_1 x_2 + m_3 x_2^2 
	+ m_4 x_1 + m_5 x_2 
	\big) \Big\} \notag\\
	&\quad \times \exp \Big\{ i \big( 
	x_1 (\omega_1 b_{11} + \omega_2 b_{12}) 
	+ x_2 (\omega_1 b_{12} + \omega_2 b_{22}) 
	\big) \Big\}
\end{align*}
Let $\mathbf{k}=(k_1, k_2, k_3, k_4, k_5)$ and $\mathbf{m}=(m_1, m_2, m_3, m_4, m_5)$. Then, we have
\begin{align}\label{S2E2}
	&\mathcal{K}_{\Omega}(x_1, x_2,\omega_1, \omega_2)\notag\\	&~~~~=\frac{i\,\sqrt{\det(B)}}{2\pi}\,\mathcal{C}_{\mathbf{k}}(\omega_1,\omega_2)\,\mathcal{C}_{\mathbf{m}}(x_1,x_2)\,\exp \Big\{ i \big( x_1 (\omega_1 b_{11} + \omega_2 b_{12}) + x_2 (\omega_1 b_{12} + \omega_2 b_{22}) \big) \Big\}, 
\end{align}
where
\begin{align}\label{S2M3}
	&\mathcal{C}_{\mathbf{k}}(\omega_1, \omega_2)=\exp \Big\{ i \big(k_1 \omega_1^2 + k_2 \omega_1 \omega_2 + k_3 \omega_2^2 + k_4 \omega_1 + k_5 \omega_2 \big) \Big\} \quad\mbox{and}\notag\\
	&\mathcal{C}_{\mathbf{m}}(x_1,x_2)=\exp \Big\{ i \big(m_1 x_1^2 + m_2 x_1 x_2 + m_3 x_2^2 + m_4 x_1 + m_5 x_2 \big) \Big\}.
\end{align}

\parindent=0mm\vspace{.1in}
As evident from the formulation, the relationship between the 2D-WD and classical 2D convolution can be elegantly expressed through a unified framework that highlights their structural similarities and operational interplay\cite{Debnath1,Feng}:
\begin{align}\label{S2E3}
	\mathcal{W}^{\Omega}_f\left(\frac{x_1}{2}, \frac{x_2}{2}, \omega_1, \omega_2 \right)  &=4 \left(\left[ f({x_1,x_2})\, e^{-i\,(\omega_1 x_1 + \omega_2 x_2)} \right] * \left[ \overline{f(x_1,x_2)}\, e^{i\,(\omega_1 x_1 + \omega_2 x_2) }\right]\right),
\end{align}
where $*$ denotes the standard two-dimensional convolution, defined as
\begin{align}\label{S2E4}
	(f * g)(x_1,x_2) &= \int_{\mathbb{R}^2} f(\tau,\eta)\,g(x_1-\tau,x_2-\eta)\,d\tau d\eta,
\end{align}
for any functions $f,g\in L^2(\mathbb{R}^2) $.

\parindent=8mm\vspace{.1in}
By replacing the two-dimensional Fourier kernel  $e^{-i\,(\omega_1 x_1 + \omega_2 x_2)}$  with $\mathcal{K}_{\Omega}(x_1, x_2,\omega_1, \omega_2)$ and $e^{i\,(\omega_1 x_1 + \omega_2 x_2)}$ with $\overline{\mathcal{K}_{\Omega}(\omega_1, \omega_2,x_1,x_2)}$ defined  in \eqref{S2E3}, we
obtain the 2D-WD in the quadratic-phase domain. This leads to the formal definition of the 2D-NSQPFT as:
\begin{align}\label{S2E5}
	\mathcal{W}^{\Omega}_f\left(\frac{x_1}{2}, \frac{x_2}{2}, \omega_1, \omega_2 \right) 
	&= 4 \big[ f(x_1, x_2) \mathcal{K}_{\Omega}(x_1, x_2,\omega_1, \omega_2)\big]  * \overline{\big[ f(x_1, x_2)\mathcal{K}_{\Omega}(\omega_1, \omega_2,x_1, x_2) \big]}. 
\end{align}
Utilizing \eqref{S2E4} and \eqref{S2E5}, and applying a change of variables, we obtain
\begin{align}\label{S2E6}
	\mathcal{W}_f^{\Omega}(x_1, x_2,\omega_1, \omega_2 ) 
	&= \int_{\mathbb{R}^2} f\left(x_1 + \frac{\tau}{2}, x_2 + \frac{\eta}{2} \right) \,
	\overline{f\left(x_1 - \frac{\tau}{2}, x_2 - \frac{\eta}{2} \right)}\notag \\
	&\quad \times \mathcal{K}_{\Omega}\left(x_1 + \frac{\tau}{2}, x_2 + \frac{\eta}{2}, \omega_1, \omega_2  \right) 
	\overline{\mathcal{K}_{\Omega}\left(\omega_1, \omega_2, x_1 - \frac{\tau}{2}, x_2 - \frac{\eta}{2} \right) }\, d\tau \, d\eta.
\end{align}
 By applying equation~\eqref{S2E2} and utilizing the identity
 \begin{align*}
 \mathcal{C}_{\mathbf{k}}(\omega_1, \omega_2) \,\overline{\mathcal{C}_{\mathbf{m}}(\omega_1, \omega_2)} =\mathcal{C}_{\mathbf{k-m}}(\omega_1, \omega_2),
 \end{align*}
 We simplify the expression to obtain the following result:
\begin{align*}
	\mathcal{W}_f^{\Omega}(x_1, x_2, \omega_1, \omega_2) 
	&= \frac{|\det B|\,\mathcal{C}_{\mathbf{k-m}}(\omega_1, \omega_2) }{(2 \pi)^2}
	\int_{\mathbb{R}^2} f\left( x_1 + \frac{\tau}{2}, x_2 + \frac{\eta}{2} \right) 
	\overline{f\left( x_1 - \frac{\tau}{2}, x_2 - \frac{\eta}{2} \right)} \\
	&\quad \times \mathcal{C}_{\mathbf{m}}\left( x_1 + \frac{\tau}{2}, x_2 + \frac{\eta}{2} \right)
	\overline{\mathcal{C}_{\mathbf{k}}\left( x_1 - \frac{\tau}{2}, x_2 - \frac{\eta}{2} \right) }\\
	&\quad \times \exp \big\{ i \left[ \omega_1(\tau b_{11} + \eta b_{12}) + 
	\omega_2 (\tau b_{12} + \eta b_{22}) \right] \big\} \, d\tau d\eta.
\end{align*}	
Equivalently, the transform can be expressed as
\begin{align*}
	\mathcal{W}_f^{\Omega}(x_1, x_2, \omega_1, \omega_2) 
	&= \frac{|\det B|\,\mathcal{C}_{\mathbf{k-m}}(\omega_1, \omega_2) }{(2 \pi)^2}\int_{\mathbb{R}^2} f_\mathbf{m}\left( x_1 + \frac{\tau}{2}, x_2 + \frac{\eta}{2} \right) 
	\overline{f_\mathbf{k}\left( x_1 - \frac{\tau}{2}, x_2 - \frac{\eta}{2} \right) }\\
	&\quad \times \exp \big\{ i \left[ \omega_1(\tau b_{11} + \eta b_{12}) + 
	\omega_2 (\tau b_{12} + \eta b_{22}) \right] \big\} \, d\tau \,d\eta,
\end{align*}
where 
\begin{align*}
	f_\mathbf{m}({\bf t})=f({\bf t})\,\mathcal{C}_{\mathbf{m}}({\bf t}) \quad \mbox{and} \quad f_\mathbf{k}({\bf t})=f({\bf t})\,\mathcal{C}_{\mathbf{k}}({\bf t}).
\end{align*}
Introducing vector notation for compactness and clarity as
\begin{align*}
  \bf{x} = \begin{pmatrix} x_1 \\ x_2 \end{pmatrix} , \quad
\boldsymbol{\xi} = \begin{pmatrix} \tau \\ \eta \end{pmatrix},\quad \mbox{and} \quad
{ \boldsymbol{\omega}} = \begin{pmatrix} \omega_1 \\ \omega_2 \end{pmatrix} .
\end{align*}
\begin{definition}\label{S2D1}
The 2D-NSQPWD of any function $ f \in L^2(\mathbb{R}^2)$ corresponding to  matrix tuple $\Omega$ is defined as:
\begin{align}\label{S2E7}
	\mathcal{W}_f^{\Omega}\left(\bf{x}, \boldsymbol{\omega}\right) 
	&=  \frac{|\det B|\,\mathcal{C}_{\mathbf{k-m}}({\boldsymbol{\omega}}) }{(2 \pi)^2} \int_{\mathbb{R}^2} f_\mathbf{m}\left({\bf{x}} +\tfrac{1}{2} {\boldsymbol{\xi}} \right) 
	\overline{f_\mathbf{k}\left( {\bf{x}} - \tfrac{1}{2} {\boldsymbol{\xi}} \right)}\,\exp\big\{i\, {\boldsymbol{\omega}^T}B {\boldsymbol{\xi}} \big\} \, d\boldsymbol{\xi},
\end{align}	
where $f_\mathbf{m}({\bf t})=f({\bf t})\,\mathcal{C}_{\mathbf{m}}({\bf t})$ and $ f_\mathbf{k}({\bf t})=f({\bf t})\,\mathcal{C}_{\mathbf{k}}({\bf t})$.

\end{definition}
\parindent=8mm\vspace{.1in}
We now explore several notable special cases of equation \eqref{S2E7}, as outlined below.
	\begin{remark}\label{S2R1}
	(i) To facilitate frequency-domain analysis, the Wigner distribution can be rewritten using the cross-distribution $\mathcal{W}_{f_\mathbf{m},f_\mathbf{k}}^{\Omega}$ 
	as follows:
	 \begin{align}\label{S2E8}
		\mathcal{W}_f^{\Omega}\left(\bf{x}, \boldsymbol{\omega}\right) &=\frac{|\det B|\,\mathcal{C}_{\mathbf{k-m}}({\boldsymbol{\omega}}) }{(2 \pi)^2}\,\mathcal{W}_{f_\mathbf{m},f_\mathbf{k}}^{\Omega}({\bf{x}}, -B{\boldsymbol{\omega}}). 
	\end{align}
	When $ B = -I $, the Wigner and cross-Wigner distributions are directly proportional at the same frequency, while $ B = I$ introduces a frequency inversion, linking them through spectral symmetry.
	
	\parindent=0mm\vspace{.0in}
	(ii) By assigning the matrix parameters as $A =\mathbf{0}_{2 \times 2}, C =\mathbf{0}_{2 \times 2}, D= \mathbf{0}_{2 \times 2}, E = \mathbf{0}_{2 \times 2}, \quad B = I_{2 \times 2},$
	and applying the scaling factor $ 4\pi^2$, Definition \ref{S2D1} reduces to the classical 2D-Wigner Distribution \eqref{S1E4} , confirming that the proposed formulation includes the standard 2D-WD as a special case.
	
	\parindent=0mm\vspace{.0in}
	(iii) By selecting the transformation matrices from the set
	 \begin{align}\label{S2E9}
		\Omega = \Bigg\{ & A = 
		\begin{bmatrix}
			a_{11} & 0 \\
			0 & a_{22}
		\end{bmatrix}, \,
		B = 
		\begin{bmatrix}
			b_{11} & 0\\
			0 & b_{22}
		\end{bmatrix}, \, \det(B) \ne 0, \,
		C = 
		\begin{bmatrix}
			c_{11} & 0 \\
			0 & c_{22}
		\end{bmatrix}, \,
		D = 
		\begin{bmatrix}
			d_{11} & 0 \\
			0 & d_{22}
		\end{bmatrix}, \notag\\
		& ~~E = 
		\begin{bmatrix}
			e_{11} & 0 \\
			0 & e_{22}
		\end{bmatrix}
		\Bigg\}.
	\end{align}
	The 2D-NSQPFT simplifies to  2D-separable quadratic phase transform (2D-SQPFT). Several well-known transforms arise as special cases of the two-dimensional separable quadratic phase Fourier transform (2D-SQPFT), such as the 2D-FT, 2D Fresnel transform, 2D-FrFT, 2D-LCT, and 2D scaling transform. \cite{Wei}. As a result, the 2D-SQPFT framework allows the construction of novel time–frequency distributions that generalize and extend these classical transforms.
	
	\parindent=0mm\vspace{.0in}
	(iv) Let
	\begin{align}\label{S2E10}
		\Omega_\theta = \Bigg\{ & A = 
		\begin{bmatrix}
		0 & \frac{1}{2}\cot\theta \\
		\frac{1}{2}\cot\theta & 	0
		\end{bmatrix}, \,
		B = 
		\begin{bmatrix}
			0 & -\csc\theta\\
		-\csc\theta &0
		\end{bmatrix}, \,
		C = 
		\begin{bmatrix}
			0 & \frac{1}{2}\cot\theta \\
			\frac{1}{2}\cot\theta & 	0
		\end{bmatrix},\notag\\
		&D = 
		\begin{bmatrix}
			0 & 0 \\
			0 & 0
		\end{bmatrix},\, 
		E = 
		\begin{bmatrix}
			0 & 0 \\
			0 & 0
		\end{bmatrix}
		\Bigg\}.
			\end{align}
	In this specific setting, 2D-NSQPWD effectively reduces to the Gyrator Transform\cite{Gyrator}, offering a powerful framework for various applications. Its ability to perform controlled rotations in phase space makes it particularly valuable in fields such as digital holography, image analysis, optical beam characterization, mode conversion, and quantum information processing, where precise manipulation of spatial and spectral content is essential.
	\begin{align}\label{S2E11}
		\mathcal{W}_f^{\Omega_\theta}({\bf{x}}, \boldsymbol{\omega}) = \frac{|\csc^2\theta|}{(2\pi)^2} \int_{\mathbb{R}^2} f\left( {\bf{x}} + \tfrac{1}{2} \boldsymbol{\xi} \right) \overline{f\left( {\bf{x}} - \tfrac{1}{2} \boldsymbol{\xi} \right)} \, e^{i\, {\bf{x}}^T J \boldsymbol{\xi} \cot\theta - i\, \boldsymbol{\omega}^T J \boldsymbol{\xi} \csc\theta} \, d\boldsymbol{\xi},
	\end{align}
where $	J = 
\begin{bmatrix}
	0 & 1 \\
	1 & 0
\end{bmatrix}$, this  expressions represent the 2D-WD corresponding to the Gyrator Transform. When the rotation angle is set to $\theta=\frac{\pi}{2}$, equation \eqref{S2E11} simplifies to the classical 2D-Wigner distribution given in \eqref{S1E4}.

	\parindent=0mm\vspace{.0in}
	(v) For a matrix
	\begin{align}\label{S2E10}
		\Omega_\theta = \Bigg\{ & A = 
		\begin{bmatrix}
	 \frac{1}{2}\cot\theta_1 &0 \\
		0 &  \frac{1}{2}\cot\theta_2
		\end{bmatrix}, \,
		B = 
		\begin{bmatrix}
			 -\csc\theta_1 &0\\
		0 & -\csc\theta_2
		\end{bmatrix},\,
		C = 
		\begin{bmatrix}
			 \frac{1}{2}\cot\theta_1 &0 \\
				0 &  \frac{1}{2}\cot\theta_2
				\end{bmatrix}, \notag\\
		&D = 
		\begin{bmatrix}
			0 & 0 \\
			0 & 0
		\end{bmatrix},\, 
		E = 
		\begin{bmatrix}
			0 & 0 \\
			0 & 0
		\end{bmatrix}
		\Bigg\}
			\end{align}
	with $\theta_i\neq k \pi$, for all $k\in \mathbb{Z} $ and $i\in \{1,2\}$, the 2D-NSQPWD \eqref{S2E7} simply yields the 2D-WD associated with the fractional Fourier transform: 
		\begin{align}\label{S2E13}
			\mathcal{W}_f^{\Omega_\theta}({\bf{x}}, \boldsymbol{\omega}) = \frac{|\csc\theta_1\,\csc\theta_2|}{(2\pi)^2} \int_{\mathbb{R}^2} f\left( {\bf{x}} + \tfrac{1}{2} \boldsymbol{\xi} \right) \overline{f\left( {\bf{x}} - \tfrac{1}{2} \boldsymbol{\xi} \right)} \, e^{i\, {\bf{x}}^T P \boldsymbol{\xi} - i\, \boldsymbol{\omega}^T Q \boldsymbol{\xi}} \, d\boldsymbol{\xi},
		\end{align}
		where  $	P = 
		\begin{bmatrix}
			\cot\theta_1& 0 \\
			0 &\cot\theta_2
		\end{bmatrix}$ and  $	Q = 
		\begin{bmatrix}
		\csc\theta_1& 0 \\
			0 & \csc\theta_2
		\end{bmatrix}$.
\end{remark}

\section{Properties of the 2D-NSQPWD}\label{S3}
In the next section, we explore several key properties of the 2D-NSQPWD \eqref{S2E7}, which play a significant role in analyzing cross-term behavior, time-bandwidth relationships, and spectrum analysis of signals in general.

\parindent=0mm\vspace{.1in}
\textbf{1. Conjugate-covariance property:}  The 2D-NSQPWD \eqref{S2E7} satisfies:
\begin{align}\label{S3E1}
	\overline{\mathcal{W}_f^{\Omega}\left(\bf{x}, \boldsymbol{\omega}\right)}	&=\mathcal{W}_f^{\Omega^\prime}\left(\bf{x}, \boldsymbol{\omega}\right),
\end{align}
where matrix parameter $\Omega^\prime=\{C,B,A,E,D\}$ .\\
\begin{proof} By virtue of Definition \ref{S2D1}, we can write
\begin{align*}
	\overline{\mathcal{W}_f^{\Omega}\left(\bf{x}, \boldsymbol{\omega}\right)}&= \frac{|\det B|\,\mathcal{C}_\mathbf{m-k}(\boldsymbol{\omega}) }{(2 \pi)^2}\int_{\mathbb{R}^2} \overline{f_\mathbf{m}\left( {\bf{x}}+ \tfrac{1}{2} \boldsymbol{\xi} \right)}\, 
	f_\mathbf{k}\left({\bf{x}} - \tfrac{1}{2} \boldsymbol{\xi} \right)\,\exp\left\{-i\, \boldsymbol{\omega}^T B \boldsymbol{\xi} \right\}d\boldsymbol{\xi}.
\end{align*}
Changing the variables $\boldsymbol{\xi}=-\boldsymbol{\xi_1}$ in above equation yields
\begin{align*}
	\overline{\mathcal{W}_f^{\Omega}\left(\bf{x}, \boldsymbol{\omega}\right)}&= \frac{|\det B|\,\mathcal{C}_\mathbf{m-k}(\boldsymbol{\omega}) }{(2 \pi)^2}\int_{\mathbb{R}^2} \overline{f_\mathbf{m}\left( \bf{x} - \tfrac{1}{2} \boldsymbol{\xi_1} \right)}\, 
	f_\mathbf{k}\left( \bf{x} + \tfrac{1}{2} \boldsymbol{\xi_1} \right)\,\exp\left\{i\, \boldsymbol{\omega}^T B \boldsymbol{\xi_1} \right\}d\boldsymbol{\xi_1}.
\end{align*}
Let $\Omega^\prime=\{C,B,A,E,D\}$ satisfy the kernal \eqref{S1E3}. Moreover, using \eqref{S2E1}, we obtain 
\begin{align}\label{S3E2}
	\begin{bmatrix}
		k_1^\prime \\ k_2^\prime \\ k_3^\prime \\ k_4^\prime \\ k_5^\prime
	\end{bmatrix}
	=
	\begin{bmatrix}
			c_{11} \\
			c_{12} + c_{21} \\
			c_{22} \\
			e_{11} + e_{21} \\
			e_{12} + e_{22}
		\end{bmatrix},
	\quad
	\begin{bmatrix}
		m_1^\prime \\ m_2^\prime \\ m_3^\prime \\ m_4^\prime \\ m_5^\prime
	\end{bmatrix}
	=	\begin{bmatrix}
			a_{11} \\
			a_{12} + a_{21} \\
			a_{22} \\
			d_{11} + d_{21} \\
			d_{12} + d_{22}
		\end{bmatrix}.
		\end{align}
Therefore
\begin{align}\label{S3E3}
	\begin{cases}
		\mathbf {k}^\prime=(k_1^\prime,k_2^\prime,k_3^\prime,k_4^\prime,k_5^\prime)=(m_1,m_2,m_3,m_4,m_5)=\mathbf{m}\\
		\mathbf{m}^\prime=(m_1^\prime,m_2^\prime,m_3^\prime,m_4^\prime,m_5^\prime)=(k_1,k_2,k_3,k_4,k_5)=\mathbf{k} .
	\end{cases}
\end{align}
Consequently, the required relation is expressed as
\begin{align*}
	\overline{\mathcal{W}_f^{\Omega}\left(\bf{x}, \boldsymbol{\omega}\right)}&=\frac{|\det B|\,\mathcal{C}_{\mathbf{k}^\prime-\mathbf{m}^\prime}(\boldsymbol{\omega}) }{(2 \pi)^2}\int_{\mathbb{R}^2}  f_{\mathbf{m}^\prime}\left( {\bf{x} }+ \tfrac{1}{2} \boldsymbol{\xi_1} \right)\,\overline{f_{\mathbf{k}^\prime}\left( {\bf{x} } - \tfrac{1}{2} \boldsymbol{\xi_1} \right)}\, 
	\exp\left\{i\, \boldsymbol{\omega}^T B \boldsymbol{\xi_1} \right\}d\boldsymbol{\xi_1}\notag\\
	&=\mathcal{W}_f^{\Omega^\prime}\left(\bf{x}, \boldsymbol{\omega}\right).
\end{align*}
This concludes the proof, with $\Omega^\prime$ prescribed in \eqref{S3E1}.
\end{proof}

\textbf{2. Symmetry-conjugation properties:} The 2D-NSQPWD corresponding to the time-reversed signal $\check{f}(\bf x) = f(-\bf x)$  can be expressed as:
\begin{align}\label{S3E4}
	\mathcal{W}_{\check{f}}^{\Omega}\left(\bf{x}, \boldsymbol{\omega}\right)&=\mathcal{W}_{f}^{\Omega^{\prime\prime}}(\bf{-x}, \boldsymbol{\omega}),
\end{align}
where  $\Omega^{\prime\prime} = \{A, -B, C, D, E\}$.

\begin{proof}
	Invoking Definition \ref{S2D1}, we have
	\begin{align*}
		\mathcal{W}_{\check{f}}^{\Omega}\left(\bf{x}, \boldsymbol{\omega}\right) 
		&= \frac{|\det B|\,\mathcal{C}_{\mathbf{k-m}}(\boldsymbol{\omega}) }{(2 \pi)^2}\int_{\mathbb{R}^2} {\check{f}}_m\left({\bf{x} } + \tfrac{1}{2} \boldsymbol{\xi} \right) 
		\overline{{\check{f}}_k\left( {\bf{x} } - \tfrac{1}{2} \boldsymbol{\xi} \right)}\,\exp\left\{i\, \boldsymbol{\omega}^T B \boldsymbol{\xi} \right\} d\boldsymbol{\xi}\notag
		\\
		&= \frac{|\det B|\,\mathcal{C}_{\mathbf{k-m}}(\boldsymbol{\omega}) }{(2 \pi)^2}\int_{\mathbb{R}^2} f_\mathbf{m}\left( {\bf{-x} }- \tfrac{1}{2} \boldsymbol{\xi} \right) 
		\overline{f_\mathbf{k}\left( {\bf{-x} } + \tfrac{1}{2} \boldsymbol{\xi} \right)}\,\exp\left\{i\, \boldsymbol{\omega}^T B \boldsymbol{\xi} \right\} d\boldsymbol{\xi}\notag
	\end{align*}
	Changing the variables $\boldsymbol{\zeta}=-\boldsymbol{\xi}$ in above equation gives 
	\begin{align*}	
		\mathcal{W}_{\check{f}}^{\Omega}\left(\bf{x}, \boldsymbol{\omega}\right) 	&= \frac{|\det B|\,\mathcal{C}_{\mathbf{k-m}}(\boldsymbol{\omega}) }{(2 \pi)^2}\int_{\mathbb{R}^2} f_\mathbf{m}\left( {\bf{-x} }+ \tfrac{1}{2} \boldsymbol{\zeta} \right) \overline{f_\mathbf{k}\left( {\bf{-x}}  - \tfrac{1}{2} \boldsymbol{\zeta} \right)}\,\exp\left\{i\, \boldsymbol{\omega}^T (-B) \boldsymbol{\zeta} \right\} d\boldsymbol{\zeta}\notag
		\\
		&=\mathcal{W}_{f}^{\Omega^{\prime\prime}}(\bf{-x}, \boldsymbol{\omega}),
	\end{align*}
This concludes the proof, with $\Omega^{\prime\prime}$ prescribed in \eqref{S3E4}. 
\end{proof}
\textbf{3. Marginal properties:} The time and frequency marginal properties of the 2D-NSQPWD can be
expressed as:  
\begin{align}\label{S3E5}
	\int_{\mathbb{R}^2}\mathcal{W}_{f}^{\Omega}\left(\bf{x}, \boldsymbol{\omega}\right)\,d{\bf{x}}&=\mathcal{Q}_{\Omega}[f](\boldsymbol{\omega})\, \mathcal{Q}^{-1}_{\Omega}\big[\overline{f}\big](\boldsymbol{\omega}).
	\end{align}
	and
	\begin{align}
		|f({\bf x})|^2&=\frac{\mathcal{C}_\mathbf{k-m}({\bf x})\,\mathcal{C}_{\mathbf{m-k}}\left(\tfrac{\boldsymbol{\omega}}{B}\right)}{|\det B|\,}\int_{\mathbb{R}^2} \mathcal{W}_{f}^{\Omega}\left(\bf{x}, \boldsymbol{\tfrac{\omega}{B}}\right)d{\boldsymbol{\omega}}.
	\end{align}
\begin{proof} We proceed in a similar way with the aid of definition to obtain
	\begin{align*}
		\int_{\mathbb{R}^2}\mathcal{W}_{f}^{\Omega}\left(\bf{x}, \boldsymbol{\omega}\right)\,d{\bf{x}}&=\frac{|\det B|\,\mathcal{C}_{\mathbf{k-m}}(\boldsymbol{\omega}) }{(2 \pi)^2}\notag\\
		&\qquad\times \int_{\mathbb{R}^2} \int_{\mathbb{R}^2} f_\mathbf{m}\left( {\bf{x}} + \tfrac{1}{2} \boldsymbol{\xi} \right)\overline{f_\mathbf{k}\left( {\bf{x}} - \tfrac{1}{2} \boldsymbol{\xi} \right)}\,\exp\left\{i\, \boldsymbol{\omega}^T B \boldsymbol{\xi} \right\} d\boldsymbol{\xi}\,d{\bf{x}}	
	\end{align*}
	Next, let ${\bf \tilde{x}} = \left( {\bf{x}} + \tfrac{1}{2} \boldsymbol{\xi} \right)$ and ${\bf \bar{x}} = \left( {\bf{x}} - \tfrac{1}{2} \boldsymbol{\xi} \right)$. Then $\boldsymbol{\xi} = ( {\bf \tilde{x}}- {\bf \bar{x}}) $, where ${\bf \tilde{x}}= (\tilde{x_1},{\tilde{x_2}})^T$ and ${\bar{x}}=({\bar {x}_1},{\bar{x_2}})^T$. With these definitions, we obtain 
	\begin{align*}
		\int_{\mathbb{R}^2}\mathcal{W}_{f}^{\Omega}\left(\bf{x}, \boldsymbol{\omega}\right)\,d{\bf{x}}&=\frac{|\det B|\,\mathcal{C}_{\mathbf{k-m}}(\boldsymbol{\omega}) }{(2 \pi)^2}\int_{\mathbb{R}^2} \int_{\mathbb{R}^2} f_\mathbf{m}\left( {\bf{\tilde x}}  \right)\overline{f_\mathbf{k}\left( {\bf{\bar x}} \right)}\,\exp\left\{i\, \boldsymbol{\omega}^T B \boldsymbol{\xi} \right\} d{\bf \bar x}\,d{\bf{\tilde x}}	\notag
		\\
		&=\frac{i\,\mathcal{C}_{\mathbf{k}}(\boldsymbol{\omega})\,\sqrt{\det(B)}}{2\pi}\left[\int_{\mathbb{R}^2} f\left( {\bf{\tilde x}}\right)\,\mathcal{C}_{\mathbf{m}}({\bf{\tilde x}})\,\exp\left\{i\, \boldsymbol{\omega}^T B {\bf{\tilde x}} \right\}\,d{\bf{\tilde x}}\right]\notag\\
		&\qquad\times \frac{-i\,\mathcal{C}_{-m}(\boldsymbol{\omega})\,\overline{\sqrt{\det(B)}}}{2\pi}\left[\int_{\mathbb{R}^2} \overline{f\left( {\bf{\bar x}}\right)} \,\mathcal{C}_{-k}({\bf{\bar x}})\,\exp\left\{-i\, \boldsymbol{\omega}^T B {\bf{\bar x}}\right\}\,d{\bf{\bar x}}\right]\notag\\
		&=\mathcal{Q}_{\Omega}[f](\boldsymbol{\omega})\, \mathcal{Q}^{-1}_{\Omega}\big[\overline{f}\big](\boldsymbol{\omega}).
	\end{align*}
	Moreover, we have
	\begin{align*}
       \mathcal{W}_{f}^{\Omega}\left(\bf{x}, \boldsymbol{\omega}\right)  &=\frac{|\det B|\,\mathcal{C}_{\mathbf{k-m}}({\boldsymbol{\omega}}) }{(2 \pi)^2}\int_{\mathbb{R}^2} f_\mathbf{m}\left({\bf{x}} +\tfrac{1}{2} {\boldsymbol{\xi}} \right) 
		\overline{f_\mathbf{k}\left( {\bf{x}} - \tfrac{1}{2} {\boldsymbol{\xi}} \right)}\,\exp\left\{i\, {(B\boldsymbol{\omega})^T} {\boldsymbol{\xi}} \right\} d{\boldsymbol{\xi}}\\
		&=\frac{|\det B|\,\mathcal{C}_{\mathbf{k-m}}({\boldsymbol{\omega}}) }{2 \pi}\,\mathcal{F}\left[f_\mathbf{m}\left({\bf{x}} +\tfrac{1}{2} {\boldsymbol{\xi}} \right) 
				\overline{f_\mathbf{k}\left( {\bf{x}} - \tfrac{1}{2} {\boldsymbol{\xi}} \right)}\right]\left(B\boldsymbol{\omega}\right)
	\end{align*}
	where $\mathcal{F}$ represents the classical Fourier transform. Now, applying the inverse Fourier transform
	in the above expression yields  
	\begin{align}\label{S3E7}
		f_\mathbf{m}\left({\bf{x}} +\tfrac{1}{2} {\boldsymbol{\xi}} \right) 
		\overline{f_\mathbf{k}\left( {\bf{x}} - \tfrac{1}{2} {\boldsymbol{\xi}} \right)}&=\frac{1}{|\det B|\,\mathcal{C}_{\mathbf{k-m}}\left(\tfrac{\boldsymbol{\omega}}{B}\right) }\int_{\mathbb{R}^2} \mathcal{W}_{f}^{\Omega}\left({\bf{x}}, \tfrac{\boldsymbol{\omega}}{B}\right)\exp\left\{-i\, {\boldsymbol{\omega}^T} {\boldsymbol{\xi}} \right\} d{\boldsymbol{\omega}}
		\end{align}
		Substituting $\boldsymbol{\xi}=0$ in \eqref{S3E7}, we obtain
			\begin{align*}
			f_\mathbf{m}\left({\bf x}\right)\overline{f_\mathbf{k}\left({\bf x}\right)}&=\frac{1}{|\det B|\,\mathcal{C}_{\mathbf{k-m}}\left(\tfrac{\boldsymbol{\omega}}{B}\right) }\int_{\mathbb{R}^2} \mathcal{W}_{f}^{\Omega}\left(\bf{x}, \boldsymbol{\tfrac{\omega}{B}}\right)d{\boldsymbol{\omega}}
				\end{align*}
				which can be expressed as:
					\begin{align*}
					|f({\bf x})|^2&=\frac{\mathcal{C}_\mathbf{k-m}({\bf x})\,\mathcal{C}_{\mathbf{m-k}}\left(\tfrac{\boldsymbol{\omega}}{B}\right)}{|\det B|\,}\int_{\mathbb{R}^2} \mathcal{W}_{f}^{\Omega}\left(\bf{x}, \boldsymbol{\tfrac{\omega}{B}}\right)d{\boldsymbol{\omega}}.
					\end{align*}
	This concludes the proof.
\end{proof}
\textbf{4. Moyal's formula:} Let $\mathcal{W}_{f}^{\Omega}\left(\bf{x}, \boldsymbol{\omega}\right)$ and $\mathcal{W}_{g}^{\Omega}\left(\bf{x}, \boldsymbol{\omega}\right) $ be the LCSWD of the functions $f$ and $g$, respectively. then, the following relation holds:
\begin{align}\label{S3E6}
	\int_{\mathbb{R}^4}\mathcal{W}_{f}^{\Omega}\left(\bf{x}, \boldsymbol{\omega}\right)\,\overline{\mathcal{W}_{g}^{\Omega}\left(\bf{x}, \boldsymbol{\omega}\right)}\,d{\bf{x}}\,d{\boldsymbol{\omega}}&=\frac{|\det B|   }{(2 \pi)^2}| \, \langle f,g\rangle|^2.
\end{align}	
The usual inner product in $ L^2(\mathbb{R}^2) $, denoted by $ \langle f, g \rangle $, is given by
\begin{align*}
	\langle f, g \rangle = \int_{\mathbb{R}^2} f({\bf x)}\,\overline{ g(\bf x)}\, d{\bf x}.
\end{align*}
\begin{proof} By virtue of Definition \ref{S2D1}, we have
	\begin{align}
		&\int_{\mathbb{R}^2}\int_{\mathbb{R}^2}\mathcal{W}_{f}^{\Omega}\left(\bf{x}, \boldsymbol{\omega}\right)\,\overline{\mathcal{W}_{g}^{\Omega}\left(\bf{x}, \boldsymbol{\omega}\right)}\,d{\bf{x}}\,d{\boldsymbol{\omega}}\notag\\
		&=\left(\frac{|\det B  |  }{(2 \pi)^2} \right)^2\int_{\mathbb{R}^2}\int_{\mathbb{R}^2}\int_{\mathbb{R}^2} \int_{\mathbb{R}^2}f_\mathbf{m}\left( {\bf{x}} + \tfrac{1}{2} {\boldsymbol{\xi}} \right) 
		\overline{f_\mathbf{k}\left( {\bf{x}} - \tfrac{1}{2} {\boldsymbol{\xi}} \right)}\notag\\
		&\qquad\times \overline{g_m\left( \bf{x} + \tfrac{1}{2} \boldsymbol{\zeta} \right)}\, 
		g_k\left( \bf{x} - \tfrac{1}{2} \boldsymbol{\zeta} \right)\,\exp\left\{i\, \boldsymbol{\omega}^T B {\boldsymbol{\xi}}\right\}\,\exp\left\{-i\, \boldsymbol{\omega}^T B \boldsymbol{\zeta} \right\} \,d\boldsymbol{\xi}\, d\boldsymbol{\zeta}\,d{\bf{x}}\,d{\boldsymbol{\omega}}\notag
	\end{align}
			where $ \boldsymbol{\zeta} =(\sigma, \lambda)$ and $d{\boldsymbol{\zeta}=d{\sigma}}\,d{\lambda}$
		\begin{align}
		&\int_{\mathbb{R}^2}\int_{\mathbb{R}^2}\mathcal{W}_{f}^{\Omega}\left(\bf{x}, \boldsymbol{\omega}\right)\,\overline{\mathcal{W}_{g}^{\Omega}\left(\bf{x}, \boldsymbol{\omega}\right)}\,d{\bf{x}}\,d{\boldsymbol{\omega}}\notag\\
			&=\left(\frac{|\det B  | }{(2 \pi)^2} \right)^2\int_{\mathbb{R}^2}\int_{\mathbb{R}^2} \int_{\mathbb{R}^2}f_\mathbf{m}\left( {\bf{x}} + \tfrac{1}{2} {\boldsymbol{\xi}} \right) 
		\overline{f_\mathbf{k}\left( {\bf{x}} - \tfrac{1}{2} {\boldsymbol{\xi}} \right)}\notag\\
		&\qquad\times \overline{g_m\left( \bf{x} + \tfrac{1}{2} \boldsymbol{\zeta} \right)}\, 
		g_k\left( \bf{x} - \tfrac{1}{2} \boldsymbol{\zeta} \right)\,\left[\int_{\mathbb{R}^2}\exp\left\{i\, \boldsymbol{\omega}^T B ({\boldsymbol{\xi}-\boldsymbol{\zeta}})\right\}\,d{\boldsymbol{\omega}}\right]\,d\boldsymbol{\xi}\, d\boldsymbol{\zeta}\,d{\bf{x}}\notag
		\\     
		&=\frac{|\det B|   }{(2 \pi)^2} \int_{\mathbb{R}^2}\int_{\mathbb{R}^2} \int_{\mathbb{R}^2}f_\mathbf{m}\left( {\bf{x}} + \tfrac{1}{2} {\boldsymbol{\xi}} \right) 
		\overline{f_\mathbf{k}\left( {\bf{x}} - \tfrac{1}{2} {\boldsymbol{\xi}} \right)}\,\overline{g_m\left( \bf{x} + \tfrac{1}{2} \boldsymbol{\zeta} \right)}\, 
		g_k\left( \bf{x} - \tfrac{1}{2} \boldsymbol{\zeta} \right)\notag\\
		&\qquad\times \,\delta(\boldsymbol{\xi}-\boldsymbol{\zeta})\,d\boldsymbol{\xi}\, d\boldsymbol{\zeta}\,d{\bf{x}}\notag	
		\\
		&=\frac{|\det B|   }{(2 \pi)^2} \int_{\mathbb{R}^2}\int_{\mathbb{R}^2} f_\mathbf{m}\left( {\bf{x}} + \tfrac{1}{2} {\boldsymbol{\xi}} \right) 
		\overline{f_\mathbf{k}\left( {\bf{x}} - \tfrac{1}{2} {\boldsymbol{\xi}} \right)}\,\overline{g_m\left( \bf{x} + \tfrac{1}{2} \boldsymbol{\zeta} \right)}\, 
		g_k\left( \bf{x} - \tfrac{1}{2} \boldsymbol{\zeta} \right)\notag\\
		&\qquad\times \,\left[\int_{\mathbb{R}^2}\delta(\boldsymbol{\xi}-\boldsymbol{\zeta})\,d\boldsymbol{\zeta}\right]d\boldsymbol{\xi}\,d{\bf{x}}\notag
		\\
		&=\frac{|\det B|   }{(2 \pi)^2} \int_{\mathbb{R}^2}\int_{\mathbb{R}^2}f_\mathbf{m}\left( {\bf{x}} + \tfrac{1}{2} {\boldsymbol{\xi}} \right) 
		\overline{f_\mathbf{k}\left( {\bf{x}} - \tfrac{1}{2} {\boldsymbol{\xi}} \right)}\,\overline{g_m\left( {\bf{x}} + \tfrac{1}{2} {\boldsymbol{\xi}} \right)}\, 
		g_k\left( {\bf{x}} - \tfrac{1}{2} {\boldsymbol{\xi}} \right) d\boldsymbol{\xi}\,d{\bf{x}}.\notag
	\end{align}
	Making the substitution $ \tilde{{\bf x}} =  {\bf x} + \tfrac{1}{2} \boldsymbol{\xi} $ and $ \bar{\bf x} =  {\bf x} - \tfrac{1}{2} \boldsymbol{\xi} $, we get 
	\begin{align*}
    	\int_{\mathbb{R}^2}\int_{\mathbb{R}^2}\mathcal{W}_{f}^{\Omega}\left(\bf{x}, \boldsymbol{\omega}\right)\,\overline{\mathcal{W}_{g}^{\Omega}\left(\bf{x}, \boldsymbol{\omega}\right)}\,d{\bf{x}}\,d{\boldsymbol{\omega}}&=\frac{|\det B|   }{(2 \pi)^2}\left[ \int_{\mathbb{R}^2}f_\mathbf{m}(\tilde{\bf x})\overline{g_m (\tilde{\bf x})}\,d{\tilde{\bf x}}\right]\left[\int_{\mathbb{R}^2} \overline{f_\mathbf{k}(\bar{\bf x})}\,	g_k(\bar{\bf x}) \,d{\bar{\bf x}}\right]\notag
		\\
		&=\frac{|\det B|   }{(2 \pi)^2}\left[ \int_{\mathbb{R}^2}f(\tilde{\bf x})\overline{g (\tilde{\bf x})}\,d{\tilde{\bf x}}\right]\left[\int_{\mathbb{R}^2} \overline{f(\bar{\bf x})}\,	g(\bar{\bf x}) \,d{\bar{\bf x}}\right]\notag
		\\ &=\frac{|\det B|   }{(2 \pi)^2}\,\big|\langle f,g\rangle\big|^2.
	\end{align*}
	This concludes the proof.
\end{proof}
	\begin{remark}\label{S2R2}
	For the special case when $f = g$, Moyal’s formula leads to an energy conservation relation for the 2D-NSQPWD \eqref{S2E7}, expressed as follows:
	\begin{align}\label{S3E8}
		\int_{\mathbb{R}^4}\big|\mathcal{W}_{f}^{\Omega}({\bf x},\boldsymbol{\omega})\big|^2
		\,d{\bf x}\,d{\boldsymbol{\omega}}
		&= \frac{|\det B|}{(2\pi)^2}\,\lVert f \rVert_{L^2(\mathbb{R}^2)}^4.
		\end{align}	
	Thus the $L^2$--energy of the 2D-NSQPWD is proportional to the fourth power of the signal energy, with proportionality constant $|\det B|/(2\pi)^2$.
	\end{remark}
\textbf{5. Time-shift property:} The 2D-NSQPWD \eqref{S2E7} of a time-shifted signal  $\check{f}({\bf x})=f({\bf x}-{\bf x_0})$  is given by
\begin{align}
	\mathcal{W}_{\check{f}}^{\Omega}\left(\bf{x}, \boldsymbol{\omega}\right) &=\nabla_1({\bf x},\boldsymbol{\omega})\,\mathcal{W}_{f}^{\Omega}(\bf{x-x_0}, \boldsymbol{\omega+\rho}),
\end{align}
where the frequency–shift vector is
\begin{align*}
	\boldsymbol{\rho} =P\,{\bf x_0} ,
\end{align*}
with $\boldsymbol{\rho}=(\rho_1,\rho_2)^T, \,{\bf x_0}=(x_{1,0} , x_{2,0})^T $, we define
\begin{align*}
	P&= 
	\begin{bmatrix}
		\tilde{b}_{22}(m_1 + k_1) - \tilde{b}_{12}\left(\dfrac{m_2 + k_2}{2}\right) &
		\tilde{b}_{22}\left(\dfrac{m_2 + k_2}{2}\right) - \tilde{b}_{12}(m_3 + k_3) \\
		-\tilde{b}_{12}(m_1 + k_1) + \tilde{b}_{11}\left( \dfrac{m_2 + k_2}{2} \right)&
		-\tilde{b}_{12}\left(\dfrac{m_2 + k_2}{2}\right) + \tilde{b}_{11}(m_3 + k_3)
	\end{bmatrix}
	\end{align*}
where
\begin{align*}
	\tilde{b}_{11} = \frac{b_{11}}{\Delta}, \quad
	\tilde{b}_{12} = \frac{b_{12}}{\Delta}, \quad
	\tilde{b}_{22} = \frac{b_{22}}{\Delta},
	\quad \Delta = b_{11}b_{22} - b_{12}^2.
	\end{align*}
	Furthermore,
	\begin{align*}
	\nabla_1({\bf x},\boldsymbol{\omega})&=\mathcal{C}_\mathbf{m-k}(\boldsymbol{ \rho})\,\mathcal{C}_{\mathbf{k-m}}{\bf (x_0)}\exp\left\{ -i\,( {\boldsymbol{\omega}^T} Q {\boldsymbol{\rho}} )\right\} \exp\left\{ -i\, \left( 
	{\bf{x}}^T Q {\bf{x}_0} + 2 \lambda^T
	{\bf{x}_0 }\right) \right\}.
\end{align*}
Here, the quadratic–phase matrix $Q$ and the linear–phase vector $\lambda$ are
	\begin{align}\label{S3E9}
	Q &= 
	\begin{bmatrix}
		2(k_1 - m_1) & (k_2 - m_2) \\
		(k_2 - m_2) & 2(k_3 - m_3)
	\end{bmatrix} ,\qquad
	\lambda=
	\begin{bmatrix}
		k_4 - m_4 \\
		k_5 - m_5
	\end{bmatrix}.
	\end{align}
\begin{proof} 
	Owing to the notation $\mathcal{C}_{\mathbf{k}}(\boldsymbol{\omega})$, it is straightforward to verify that
	\begin{align*}
		&\mathcal{C}_{\mathbf{k}}(\boldsymbol{\omega})\notag\\&=\mathcal{C}_{\mathbf{k}}(\omega_1+\rho_1, \omega_2+\rho _2)\,\mathcal{C}_{-k}(\rho_1, \rho_2)\, \exp\left\{-i\,((2k_1\rho_1+k_2\rho_2 )\omega _1+(2k_3\rho_2+ k_2\rho_1)\omega_2)\right\}
	\end{align*}
	and  
	\begin{align*}
		&\mathcal{C}_{\mathbf{m}}(\boldsymbol{\omega})\notag\\
		&=\mathcal{C}_{\mathbf{m}}(\omega_1+\rho_1, \omega_2+\rho _2)\,\mathcal{C}_{-m}(\rho_1, \rho_2)\, \exp\left\{-i\left((2m_1\rho_1+m_2\rho_2) \omega _1+(2m_3\rho_2+ m_2\rho_1)\omega_2\right)\right\}.
	\end{align*}
	Now
	\begin{align}\label{S3E10}
		\mathcal{C}_{\mathbf{k-m}}(\boldsymbol{\omega})&=\mathcal{C}_{\mathbf{k-m}}(\boldsymbol{\omega}+\boldsymbol{\rho})\,\mathcal{C}_\mathbf{m-k}(\boldsymbol{\rho})\, \exp\{-i\,((2(k_1-m_1)\rho_1\notag
		\\
		&\qquad+(k_2-m_2)\rho_2 )\omega _1+(2(k_3-m_3)\rho_2+ (k_2-m_2)\rho_1)\omega_2)\}.
	\end{align}
	Now, by virtue of Definition \ref{S2D1}, we can write
	\begin{align}\label{S3E11}
		&\mathcal{W}_{\check{f}}^{\Omega}\left(\bf{x}, \boldsymbol{\omega}\right)\notag\\ 
		&=\frac{|\det B|\,\mathcal{C}_{\mathbf{k-m}}(\boldsymbol{\omega}) }{(2 \pi)^2} \int_{\mathbb{R}^2} {\check{f}}_m\left( {\bf{x}} + \tfrac{1}{2} {\boldsymbol{\xi}} \right) 
		\overline{\check{f}_k\left( {\bf{x}} - \tfrac{1}{2} {\boldsymbol{\xi}} \right)}\,\exp\left\{i\, \boldsymbol{\omega}^T B \boldsymbol{\xi} \right\} \, d\boldsymbol{\xi}\notag
		\\
		&=\frac{|\det B|\,\mathcal{C}_{\mathbf{k-m}}( \boldsymbol{\omega}) }{(2 \pi)^2} \int_{\mathbb{R}^2} f\left( \bf{x-x_0} + \tfrac{1}{2} \boldsymbol{\xi} \right)\, \mathcal{C}_{\mathbf{m}}{(\bf{x}+ \tfrac{1}{2} \boldsymbol{\xi})}\,\overline{f\left( \bf{x-x_0} - \tfrac{1}{2} \boldsymbol{\xi} \right)}\,\overline{\mathcal{C}_{\mathbf{k}}{(\bf{x}- \tfrac{1}{2} \boldsymbol{\xi})}}\notag\\
		&\qquad\times \exp\left\{i\, \boldsymbol{\omega}^T B \boldsymbol{\xi} \right\} \,d\boldsymbol{\xi}\notag
		\\
		&=  \frac{|\det B|\,\mathcal{C}_{\mathbf{k-m}}(\boldsymbol{\omega})\,\mathcal{C}_{\mathbf{k-m}}(\bf x_0) }{(2 \pi)^2} \int_{\mathbb{R}^2} f\left( \bf{x-x_0} + \tfrac{1}{2} \boldsymbol{\xi} \right)\, \mathcal{C}_{\mathbf{m}}{(\bf{x-x_0 + \tfrac{1}{2} \boldsymbol{\xi}})}\notag\\
		&\qquad\times\overline{f\left( \bf{x-x_0 + \tfrac{1}{2} \boldsymbol{\xi}} \right)}\,\overline{\mathcal{C}_{\mathbf{k}}{(\bf{x-x_0}- \tfrac{1}{2} \boldsymbol{\xi})}}\,\exp\{-i\,\big(2(k_1-m_1)x_1x_{1,0}\notag\\
		&\qquad+(k_2-m_2)(x_1 x_{2,0}+x_{1,0}x_2)+2(k_3-m_3)x_2x_{2,0+2(k_4-m_4)x_{1,0}}\notag\\
		&\qquad+2(k_5-m_5)x_{2,0}\big)\}\, \exp\left\{i\,(u_1 \tau+u_2 \eta)\right\}\,\exp\left\{i\,\boldsymbol{\omega}^T B \boldsymbol{\xi} \right\} \, d\boldsymbol{\xi},
	\end{align}
	where 
	\begin{align*}
		\begin{cases}
			(m_1 + k_1)x_{1,0} + \frac{(m_2 + k_2)}{2} x_{2,0} = u_1 \\
			\frac{(m_2 + k_2)}{2} x_{1,0} + (m_3 + k_3) x_{2,0} = u_2.
		\end{cases}
	\end{align*}
		On simplifying, we get 
		\begin{align}\label{S3E12}
			\exp\left\{i\, \boldsymbol{\omega}^T B \boldsymbol{\xi} \right\}&=\exp\left\{i\, (\boldsymbol{\omega}+{\boldsymbol{\rho}})^T B \boldsymbol{\xi} \right\} \exp\left\{-i\, ({\boldsymbol{\rho}}^T B \boldsymbol{\xi} )\right\}\notag
			\\
			&=\exp\left\{i\, (\boldsymbol{\omega}+{\boldsymbol{\rho}})^T B \boldsymbol{\xi} \right\}\exp\left\{ -i \left[ \tau (\rho_1 b_{11} + \rho_2 b_{21}) + \eta (\rho_1 b_{12} + \rho_2 b_{22}) \right] \right\}\notag
			\\
			&=\exp\left\{i\, (\boldsymbol{\omega}+{\boldsymbol{\rho}})^T B \boldsymbol{\xi} \right\}\exp\left\{ -i \,( \tau u_1 + \eta u_2) \right\},
		\end{align}
	where
	\begin{align*}
		\begin{cases}
			u_1= b_{11} \rho_1 + b_{12}  \rho_2  \\
			u_2= b_{12}  \rho_1 + b_{22} \rho_2.
		\end{cases}
	\end{align*}
	Finding $\rho_1 ,\rho_2$ gives us
	\begin{align*}
		\begin{cases}
			\rho_1 = \tilde {b}_{22} u_1-\tilde{b}_{12}  u_2\\
			\rho_2 = -\tilde{b}_{12} u_1 +\tilde{b}_{11} u_2,
		\end{cases}
	\end{align*}
	Substitute $u_1$ and $u_2$ ,we get
	\begin{align*}
			\begin{cases}
				\rho_1 = \tilde {b}_{22} \left((m_1 + k_1)x_{1,0} + \frac{(m_2 + k_2)}{2} x_{2,0}\right)-\tilde{b}_{12} 	\left(\frac{(m_2 + k_2)}{2} x_{1,0} + (m_3 + k_3) x_{2,0}\right)\\
				\rho_2 = -\tilde{b}_{12} \left((m_1 + k_1)x_{1,0} + \frac{(m_2 + k_2)}{2} x_{2,0}\right)+\tilde{b}_{11} 	\left(\frac{(m_2 + k_2)}{2} x_{1,0} + (m_3 + k_3) x_{2,0}\right).
			\end{cases}
		\end{align*}
	This system can be expressed compactly as $	\boldsymbol{\rho} =P\,{\bf x_0}$ , where 
	\begin{align*}
		P&= 
		\begin{bmatrix}
			\tilde{b}_{22}(m_1 + k_1) - \tilde{b}_{12}\left(\dfrac{m_2 + k_2}{2}\right) &
			\tilde{b}_{22}\left(\dfrac{m_2 + k_2}{2}\right) - \tilde{b}_{12}(m_3 + k_3) \\
			-\tilde{b}_{12}(m_1 + k_1) + \tilde{b}_{11}\left( \dfrac{m_2 + k_2}{2} \right)&
			-\tilde{b}_{12}\left(\dfrac{m_2 + k_2}{2}\right) + \tilde{b}_{11}(m_3 + k_3)
		\end{bmatrix}
	\end{align*} and
	\begin{align}\label{S3E13}
		\begin{cases}
			\tilde{b}_{11}= \frac{ b_{11}}{b_{11}b_{22} - b_{12}^2}\\
			\tilde{b}_{12}= \frac{ b_{12}}{b_{11}b_{22} - b_{12}^2}\\
			\tilde {b}_{22}=\frac{ b_{22}}{b_{11}b_{22} - b_{12}^2}.
		\end{cases}
	\end{align}

	 Substituting equations \eqref{S3E10} and \eqref{S3E12} into \eqref{S3E11}, we obtain
	\begin{align*}
		\mathcal{W}_{\check{f}}^{\Omega}\left(\bf{x}, \boldsymbol{\omega}\right)&=\frac{|\det B|\,\mathcal{C}_{\mathbf{k-m}}(\boldsymbol{\omega})\,\mathcal{C}_{\mathbf{k-m}}(\bf x_0) }{(2 \pi)^2} \int_{\mathbb{R}^2}  f_\mathbf{m}\left( {\bf{x-x_0}} + \tfrac{1}{2} {\boldsymbol{\xi}} \right)\overline{f_\mathbf{k}\left( {\bf{x-x_0}}+ \tfrac{1}{2} {\boldsymbol{\xi}} \right)}\\
		&\qquad\times \exp\{-i\,\big(2(k_1-m_1)x_1x_{1,0}+(k_2-m_2)(x_1 x_{2,0}+x_{1,0}x_2)+2(k_3-m_3)x_2x_{2,0}\\
		&\qquad +2(k_4-m_4)x_{1,0}+2(k_5-m_5)x_{2,0}\big)\}\, \exp\left\{i\,(u_1 \tau+u_2 \eta)\right\}\exp\left\{i\, {(\boldsymbol{\omega}+\boldsymbol{\rho})}^T B \boldsymbol{\xi} \right\}\\
		&\qquad\times\exp\left\{-i\,(u_1 \tau  +u_2 \eta ) \right\} \, d\boldsymbol{\xi}
		\\
		&=\mathcal{C}_\mathbf{m-k}({\boldsymbol{\rho}})\,\mathcal{C}_{\mathbf{k-m}}(\bf x_0)\\
		&\qquad\times\exp\{-i\,((2(k_1-m_1)\rho_1+(k_2-m_2)\rho_2 )\omega _1+(2(k_3-m_3)\rho_2+ (k_2-m_2)\rho_1)\omega_2)\} \\
		&\qquad\times \exp\{-i\,\big(2(k_1-m_1)x_1x_{1,0}+(k_2-m_2)(x_1 x_{2,0}+x_{1,0}x_2)+2(k_3-m_3)x_2x_{2,0}\\
		&\qquad+2(k_4-m_4)x_{1,0}++2(k_5-m_5)x_{2,0}\big)\}\left[\frac{|\det B|\,\mathcal{C}_{\mathbf{k-m}}(\boldsymbol{\omega}+\boldsymbol{\rho})}{(2 \pi)^2}\right. \\
		&\qquad\times\left. \int_{\mathbb{R}^2} f_\mathbf{m}\left( {\bf{x-x_0}} + \tfrac{1}{2} {\boldsymbol{\xi}} \right)\overline{f_\mathbf{k}\left( {\bf{x-x_0}} + \tfrac{1}{2} {\boldsymbol{\xi}} \right)}\exp\left\{i\, (\boldsymbol{\omega}+\boldsymbol{\rho})^T B \boldsymbol{\xi} \right\}\, d\boldsymbol{\xi}\right]
		\\
		&=\nabla_1({\bf x},\boldsymbol{\omega})\,\mathcal{W}_{f}^{\Omega}(\bf{x-x_0}, \boldsymbol{\omega}+\boldsymbol{\rho}),
	\end{align*}
	where 
	\begin{align*}
		&\nabla_1({\bf x},\boldsymbol{\omega})\notag\\
		&=\mathcal{C}_\mathbf{m-k}({\boldsymbol{\rho}})\,\mathcal{C}_{\mathbf{k-m}}(\bf x_0)\\
		&\qquad\times\exp\{-i\,((2(k_1-m_1)\rho_1+(k_2-m_2)\rho_2 )\omega _1+(2(k_3-m_3)\rho_2+ (k_2-m_2)\rho_1)\omega_2)\} \\
		&\qquad\times \exp\{-i\,\big(2(k_1-m_1)x_1x_{1,0}+(k_2-m_2)(x_1 x_{2,0}+x_{1,0}x_2)+2(k_3-m_3)x_2x_{2,0}\\
		&\qquad+2(k_4-m_4)x_{1,0}+2(k_5-m_5)x_{2,0}\big)\}\\
		&=\mathcal{C}_\mathbf{m-k}(\boldsymbol{ \rho})\,\mathcal{C}_{\mathbf{k-m}}{\bf (x_0)}\exp\left\{ -i\,( {\boldsymbol{\omega}^T} Q {\boldsymbol{\rho}} )\right\} \exp\left\{ -i\, \left( 
			{\bf{x}}^T Q {\bf{x}_0} + 2 \lambda^T
			{\bf{x}_0 }\right) \right\}.
	\end{align*}
	This concludes the proof, with $\mathcal{Q}$ prescribed in  \eqref{S3E9}.
\end{proof}
\textbf{6. Frequency-shift property:} The 2D-NSQPWD \eqref{S2E7} of a modulated signal $	\hat f({\bf x})=f(\bf x)\,\exp\left\{i\,({\boldsymbol{\omega_0}^T}{\bf x})\right\}$ is given by 
\begin{align}\label{S3E14}
	\mathcal{W}_{\hat{f}}^{\Omega}\left(\bf{x}, \boldsymbol{\omega}\right)&=\mathcal{C}_\mathbf{m-k}{(\bf t)}\,\exp\left\{ -i\,( {\boldsymbol{\omega}^T} Q t)\right\} \,\mathcal{W}_f^{\Omega}(\bf{x},\boldsymbol{\omega}+ {\bf t}),
\end{align}
where  
	\begin{align*}
	Q &= 
	\begin{bmatrix}
		2(k_1 - m_1) & (k_2 - m_2) \\
		(k_2 - m_2) & 2(k_3 - m_3)
	\end{bmatrix}.
	\end{align*}
\begin{proof} By virtue of Definition \ref{S2D1}, we can write
	\begin{align}
		\mathcal{W}_{\hat{f}}^{\Omega}\left(\bf{x}, \boldsymbol{\omega}\right)
		&=\frac{|\det B|\,\mathcal{C}_{\mathbf{k-m}}(\boldsymbol{\omega}) }{(2 \pi)^2} \int_{\mathbb{R}^2} f\left( {\bf{x}} + \tfrac{1}{2} {\boldsymbol{\xi}} \right)\, \mathcal{C}_{\mathbf{m}}\left( {\bf{x}} + \tfrac{1}{2} {\boldsymbol{\xi}} \right)\,\overline{f\left( {\bf{x}} - \tfrac{1}{2} {\boldsymbol{\xi}} \right)}\,\overline{\mathcal{C}_{\mathbf{k}}{\left({\bf{x}}- \tfrac{1}{2} {\boldsymbol{\xi}}\right)}}\notag\\
		&\qquad\times \exp\left\{i\,{\boldsymbol{\omega_0}^T}{\left( {\bf{x}} + \tfrac{1}{2} {\boldsymbol{\xi}} \right)}\right\}\exp\left\{-i\,{\boldsymbol{\omega_0}^T}{\left( {\bf{x}} - \tfrac{1}{2} {\boldsymbol{\xi}} \right)}\right\} \exp\left\{i\, \boldsymbol{\omega}^T B \boldsymbol{\xi} \right\} \,d\boldsymbol{\xi}\notag
		\\
		&=\frac{|\det B|\,\mathcal{C}_{\mathbf{k-m}}(\boldsymbol{\omega}) }{(2 \pi)^2} \int_{\mathbb{R}^2} f\left( {\bf{x}} + \tfrac{1}{2} {\boldsymbol{\xi}} \right)\, \mathcal{C}_{\mathbf{m}}{\left( {\bf{x}} + \tfrac{1}{2} {\boldsymbol{\xi}} \right)}\,\overline{f\left( {\bf{x}} - \tfrac{1}{2} {\boldsymbol{\xi}} \right)}\,\overline{\mathcal{C}_{\mathbf{k}}{\left( {\bf{x}} - \tfrac{1}{2} {\boldsymbol{\xi}} \right)}}\notag\\
		&\qquad\times \exp\left\{i\,{\boldsymbol{\omega_0}^T}{\boldsymbol{\xi}}\right\} \exp\left\{i\, \boldsymbol{\omega}^T B \boldsymbol{\xi} \right\} \,d\boldsymbol{\xi}\notag
		\\
		&=\frac{|\det B|\,\mathcal{C}_{\mathbf{k-m}}(\boldsymbol{\omega}) }{(2 \pi)^2} \int_{\mathbb{R}^2} f_\mathbf{m}\left( {\bf{x}} + \tfrac{1}{2} {\boldsymbol{\xi}} \right)\,\overline{f_\mathbf{k}\left( {\bf{x}} - \tfrac{1}{2} {\boldsymbol{\xi}} \right)}\exp\left\{i\,{\boldsymbol{\omega_0}^T}{\boldsymbol{\xi}}\right\} \exp\left\{i\, \boldsymbol{\omega}^T B \boldsymbol{\xi} \right\} \,d\boldsymbol{\xi}.\notag
	\end{align}	
	We now find parameters  ${\bf t}=(t_1,t_2)^T$, which satisfy
	\begin{align*}
		\exp\left\{i\, (\boldsymbol{\omega}^T B \boldsymbol{\xi} )\right\}&=\exp\left\{i\, (\boldsymbol{\omega}+{\bf t })^T B \boldsymbol{\xi} \right\} \exp\left\{-i\, ({\bf t }^T B \boldsymbol{\xi} )\right\}
		\\
		&=\exp\left\{i\, (\boldsymbol{\omega}+{\bf t })^T B \boldsymbol{\xi} \right\}\exp\left\{ -i\, \left( \tau ( b_{11} t_1 + b_{21}t_2) + \eta ( b_{12}t_1 + b_{22}t_2) \right) \right\}
		\\
		&=\exp\left\{i\, (\boldsymbol{\omega}+{\bf t })^T B \boldsymbol{\xi} \right\}\exp\left\{ -i\, ( \tau\, \omega_{1,0}+ \eta\, \omega_{2,0}) \right\}
		\\
		&=\exp\left\{i\, (\boldsymbol{\omega}+{\bf t })^T B \boldsymbol{\xi} \right\}\exp\left\{ -i\, ( {\boldsymbol{\omega_0}^T}{\boldsymbol{\xi}}) \right\}.
	\end{align*}
	Here, $t_1 ,t_2$ satisfy
	\begin{align*}
		\begin{cases}
			b_{11} t_1 + b_{12}  t_2 = \omega_{1,0} \\
			b_{12}  t_1 + b_{22} t_2 = \omega_{2,0},
		\end{cases}
	\end{align*}
	Therefore, the above equations turn into
	\begin{align*}
		\begin{cases}
			t_1 = \tilde {b}_{22} \,\omega_{1,0}-\tilde{b}_{12}\,  \omega_{2,0}\\
			t_2 = -\tilde{b}_{12} \,\omega_{1,0} +\tilde{b}_{11} \,\omega_{2,0},
		\end{cases}
	\end{align*}
	 All $\tilde{b}_{ij}$ are defined in Equation \eqref{S3E13}. Based on equation \eqref{S3E10}, we find that
	
	\begin{align*}
		\mathcal{C}_{\mathbf{k-m}}(\boldsymbol{\omega})&=\mathcal{C}_{\mathbf{k-m}}(\boldsymbol{\omega}+{\bf t})\,\mathcal{C}_\mathbf{m-k}({\bf t})\, \exp\{-i\,((2(k_1-m_1)t_1
		\\
		&\qquad+(k_2-m_2)t_2 )\omega _1+(2(k_3-m_3)t_2+ (k_2-m_2)t_1)\omega_2)\}
	\end{align*}
	Thus, we have
	\begin{align*}
		&\mathcal{W}_{\hat{f}}^{\Omega}\left(\bf{x}, \boldsymbol{\omega}\right)\notag\\
		&=\left[\frac{|\det B|\,\mathcal{C}_\mathbf{m-k}(\bf t)\,\mathcal{C}_{\mathbf{k-m}}{ (\boldsymbol{\omega}+{\bf t}) } }{(2 \pi)^2} 
		\int_{\mathbb{R}^2} f_\mathbf{m}\left( {\bf{x}} + \tfrac{1}{2} {\boldsymbol{\xi}} \right)\,\overline{f_\mathbf{k}\left( {\bf{x}} - \tfrac{1}{2} {\boldsymbol{\xi}} \right)}\exp\left\{i\,{ (\boldsymbol{\omega}+{\bf t} })^T B \boldsymbol{\xi} \right\}d\boldsymbol{\xi}\right]\notag\\
		&\qquad\times\exp\{-i\,((2(k_1-m_1)t_1+(k_2-m_2)t_2 )\omega _1+(2(k_3-m_3)t_2+ (k_2-m_2)t_1)\omega_2)\}\notag 
		\\
		&=\mathcal{C}_\mathbf{m-k}{(\bf t)}\,\exp\left\{ -i\,( {\boldsymbol{\omega}^T} Q t)\right\} \,\mathcal{W}_f^{\Omega}(\bf{x},\boldsymbol{\omega}+ {\bf t}).
	\end{align*}
This concludes the proof, with $\mathcal{Q}$ prescribed in  \eqref{S3E14}.
\end{proof} 
\textbf{7. Association with the STFT:} For a signal $f(\bf x)$, the the Short-Time Fourier Transform (STFT) is defined by
\begin{align}\label{S3E15}
	S_{f,g}\left({\bf x},{\boldsymbol{\omega}}\right)& = \int_{\mathbb{R}^2} f(\boldsymbol{\xi})\, g({\boldsymbol{\xi}}-{\bf x})\,\exp\left\{-i\, {\boldsymbol{\omega}^T}{\boldsymbol{\xi}}\right\} d\boldsymbol{\xi},
\end{align}
where $g(\bf x)$ denotes the window function.

\vspace{0.1cm}
An analytical relation between the 2D-NSQPWD and the 2D-STFT is given by \begin{align}\label{S3E16}
	\mathcal{W}_{f}^{\Omega}\left(\frac{{\bf x}}{2},\tilde{B}\boldsymbol{\omega}\right)&=\nabla_2({\bf x}, {\boldsymbol{\omega}})\,S_{f,g}\left({\bf x},\boldsymbol{\omega} \right),
\end{align}
where 
\begin{align}
	\nabla_2(\bf x, \boldsymbol{\omega})&=\frac{|\det B|\,\mathcal{C}_{\mathbf {m}}({\bf x})\,\mathcal{C}_\mathbf{k-m}\left(\tilde{B}\boldsymbol{\omega}\right)}{(2\pi)^2},\notag
\end{align}
and
\begin{align*}
	g({\bf x})&=\overline {f_\mathbf{k}\left(-{\bf x}\right)}\,\mathcal{C}_{\mathbf {m}}({\bf x})\,\exp\left\{
						- i\,{\bf{x}}^{T}
						\begin{bmatrix}
						-4m_1 & -2m_2 \\
						-2m_2 & -4m_3
						\end{bmatrix}
						\bf{x}
						\right\}\exp\left\{i\, \boldsymbol{\omega}^T B{\bf{x}} \right\}. 
\end{align*}
\begin{proof} Defining $\tilde{\bf x}=(\tilde{x_1},\tilde{x_2})^T$ with  $\Delta\tilde{\bf{x}}=2(\tilde{\bf x}-{\bf x})$, and applying the substitution $ \tilde{\bf x} = {\bf x} + \tfrac{1}{2}\boldsymbol{\xi}$, relations \eqref{S2E7} becomes
\begin{align}\label{S3E17}
\mathcal{W}_{f}^{\Omega}\left(\bf{x}, \boldsymbol{\omega}\right)
&=\frac{|\det B|\, \mathcal{C}_{\mathbf{k-m}}(\boldsymbol{\omega})}{(2\pi)^2} 
\int_{\mathbb{R}^2} f_\mathbf{m}\left(\tilde{\bf{x}}\right)\,
\overline{f_\mathbf{k}\left( 2\bf{x} - \tilde{\bf{x}} \right)}
\exp\left\{i\, \boldsymbol{\omega}^T B\Delta\tilde{\bf{x}} \right\}d\tilde{\bf{x}}\notag\\
&= \frac{|\det B|\, \mathcal{C}_{\mathbf{k-m}}(\boldsymbol{\omega})}{(2\pi)^2} 
\int_{\mathbb{R}^2} f(\tilde{\bf{x}})\,\mathcal{C}_{\mathbf{m}}(\tilde{\bf{x}})\,
\overline{f_\mathbf{k}\left( 2\bf{x} - \tilde{\bf{x}} \right)}
\exp\left\{i\, \boldsymbol{\omega}^T B\Delta\tilde{\bf{x}} \right\}d\tilde{\bf{x}} .
\end{align} 
Using equation \eqref{S2M3}, we obtain
	\begin{align}
		&\mathcal{C}_{\mathbf {m}}(\tilde{x}_1 - 2x_1, \tilde{x}_2 - 2x_2)\notag\\
		&= \exp \Big\{ i\, \big(
		m_1 \left( \tilde{x}_1^2 - 4\tilde{x}_1 x_1 + 4x_1^2 \right)
		+ m_2 \left( \tilde{x}_1 \tilde{x}_2 - 2\tilde{x}_1 x_2 - 2x_1 \tilde{x}_2 + 4x_1 x_2 \right) \notag\\
		&\quad + m_3 \left( \tilde{x}_2^2 - 4\tilde{x}_2 x_2 + 4x_2^2 \right)
		+ m_4 \left( \tilde{x}_1 - 2x_1 \right)
		+ m_5 \left( \tilde{x}_2 - 2x_2 \right)
		\big) \Big\}\notag
		\\
		&=\mathcal{C}_{\mathbf {m}}({2x}_1,{2x}_2)\,\mathcal{C}_{\mathbf {m}}(\tilde {x}_1,\tilde {x}_2)\notag\\
		&\quad\times \exp\left\{i\,(-8m_1 x_1^2-8m_2x_1 x_2-8m_3x_2^2-4m_4x_1-4m_5x_2)\right\}\notag\\
		&\quad\times \exp \left\{ i\, \left((\tilde{x}_1 - 2x_1)\left(- 4m_1 x_1 - 2m_2 x_2 \right)
		+ (\tilde{x}_2 - 2x_2)\left(- 4m_3 x_2 - 2m_2 x_1 \right) \right) \right\}\notag  
	\end{align}
	Simplifying, we get
	\begin{align}\label{S3E18}
	\mathcal{C}_{\mathbf{m}}(\tilde{\bf{x}})
	&= \mathcal{C}_{\mathbf{m}}(\tilde{\bf{x}} - 2{\bf{x}})\,
	   \mathcal{C}_{-\mathbf{m}}(2{\bf{x}})\,
	   \mathcal{C}_{\mathbf{m}}^{2}(2{\bf{x}}) \notag\\
	&\quad\times \exp\left\{
	- i\,(\tilde{\bf{x}} - 2\bf{x})^{T}
	\begin{bmatrix}
	-4m_1 & -2m_2 \\
	-2m_2 & -4m_3
	\end{bmatrix}
	\bf{x}
	\right\}
	\end{align}
Substituting equation~\eqref{S3E18} into equation~\eqref{S3E17} yields
	\begin{align*}
		&\mathcal{W}_{f}^{\Omega}\left(\bf{x}, \boldsymbol{\omega}\right)\,\mathcal{C}_\mathbf{m-k}(\boldsymbol{\omega})\notag\\
		&= \frac{|\det B|\,\mathcal{C}_{\mathbf {m}}(2{\bf x})}{(2\pi)^2} 
		\int_{\mathbb{R}^2} 
	   f\left(\tilde{\bf{x}}\right)\,
	   \overline{f_\mathbf{k}\left( 2\bf{x} - \tilde{\bf{x}} \right)}\, \mathcal{C}_{\mathbf{m}}(\tilde{\bf{x}} - 2\bf{x}) \notag\\
		&\quad\times  \exp\left\{
			- i\,(\tilde{\bf{x}} - 2\bf{x})^{T}
			\begin{bmatrix}
			-4m_1 & -2m_2 \\
			-2m_2 & -4m_3
			\end{bmatrix}
			\bf{x}
			\right\}\exp\left\{i\, \boldsymbol{\omega}^T B\Delta\tilde{\bf{x}} \right\}d\tilde{\bf{x}}\notag 
			\\
	    	&= \frac{|\det B| \,\mathcal{C}_{\mathbf {m}}(2{\bf x})}{(2\pi)^2} 
			\int_{\mathbb{R}^2}f\left(\tilde{\bf{x}}\right)\,	g(\tilde{\bf x}-2{\bf x})\,\exp\left\{i\, \boldsymbol{\omega}^T B\tilde{\bf{x}} \right\}d\tilde{\bf x},
		\end{align*}
	where 
	\begin{align}\label{S3E19}
		g(\tilde{\bf x})&=\overline {f_\mathbf{k}\left(-\tilde{\bf x}\right)}\,\mathcal{C}_{\mathbf {m}}(\tilde{\bf x})\,\exp\left\{
					- i\,\tilde{\bf{x}}^{T}
					\begin{bmatrix}
					-4m_1 & -2m_2 \\
					-2m_2 & -4m_3
					\end{bmatrix}
					\bf{x}
					\right\}\exp\left\{i\, \boldsymbol{\omega}^T B\tilde{\bf{x}} \right\}. 
	\end{align}
	This allow us to observe that
	\begin{align}
		&\mathcal{W}_{f}^{\Omega}\left(\frac{\bf{x}}{2}, \boldsymbol{\omega}\right)\mathcal{C}_\mathbf{m-k}(\boldsymbol{\omega})\notag\\
		&=\frac{|\det B|\,\mathcal{C}_{\mathbf {m}}({\bf x})}{(2\pi)^2}		\int_{\mathbb{R}^2}f\left(\tilde{\bf{x}}\right)\,	g(\tilde{\bf x}-{\bf x})\,\exp\left\{i\, \boldsymbol{\omega}^T B\tilde{\bf{x}} \right\}d\tilde{\bf x}\notag
		\\
		&=\frac{|\det B|\,\mathcal{C}_{\mathbf {m}}({\bf x})}{(2\pi)^2}\,S_{f,g}\left({\bf x},-B\boldsymbol{\omega} \right).\notag
	\end{align}
	Now, consider $\boldsymbol{\omega}_0=({\omega}_{1,0},{\omega}_{2,0})^T$. The system $B\boldsymbol{\omega} = -\boldsymbol{\omega}_0$ yields  
	\begin{align*}
	 \boldsymbol{\omega}&= \tilde{B}\,\boldsymbol{\omega}_0,
	 \quad
	 \tilde{B}=
	 \begin{bmatrix}
	 -\tilde{b}_{22} & \tilde{b}_{12} \\
	 \tilde{b}_{12} & -\tilde{b}_{11}
	 \end{bmatrix}, 
	\end{align*}
	Where all $\tilde{b}_{ij}$ are defined in equation~\eqref{S3E13}. The relation becomes
	\begin{align}
		&\mathcal{W}_{f}^{\Omega}\left(\frac{{\bf x}}{2},\tilde{B}\boldsymbol{\omega}\right)\mathcal{C}_\mathbf{m-k}\left(\tilde{B}\boldsymbol{\omega}\right)\notag\\
		&=\frac{|\det B|\,\mathcal{C}_{\mathbf {m}}({\bf x})}{(2\pi)^2}	\int_{\mathbb{R}^2}f\left(\tilde{\bf{x}}\right)\,	g(\tilde{\bf x}-{\bf x})\,\exp\left\{-i\, \boldsymbol{\omega}^T \tilde{\bf{x}} \right\}d\tilde{\bf x}\notag\\
		&=\frac{|\det B|\,\mathcal{C}_{\mathbf {m}}({\bf x})}{(2\pi)^2}\,S_{f,g}\left({\bf x},\boldsymbol{\omega} \right).\notag
	\end{align}
		This concludes the proof.
\end{proof}
\textbf{8. Dilation property:} Let $\mathcal{W}_{f}^{\Omega}\left(\bf{x}, \boldsymbol{\omega}\right)$ denote the 2D-NSQPWD of a function $f({\bf{x}})\in L^2(\mathbb{R}^2)$. Then, the following relation holds:
\begin{align}
\mathcal{W}_{D_\lambda f}^\Omega\left(\bf{x}, \boldsymbol{\omega}\right) &=\frac{1}{\lambda} \, \mathcal{W}_f^\Omega\left(\lambda \bf{x}, \frac{\boldsymbol{\omega}}{\lambda}\right).
\end{align}
where $D_\lambda f({\bf{x}}) = \sqrt{\lambda} \, f(\lambda {\bf{x}})$,~~ $\lambda\in\mathbb{R}^+$.
\begin{proof} From Definition \ref{S2D1}, we have
\begin{align*}
\mathcal{W}_{D_\lambda f}^\Omega\left(\bf{x}, \boldsymbol{\omega}\right) &= \frac{|\det B|\, \mathcal{C}_{\bf{k-m}}(\boldsymbol{\omega})}{(2\pi)^2} \int_{\mathbb{R}^2} (D_\lambda f)_{\mathbf {m}}\left({\bf x} + \tfrac{1}{2}\boldsymbol{\xi}\right) \overline{(D_\lambda f)_\mathbf{k}\left({\bf x} - \tfrac{1}{2}\boldsymbol{\xi}\right)}  \exp\left\{i\, \boldsymbol{\omega}^T B \boldsymbol{\xi} \right\} d\boldsymbol{\xi} \notag\\
&=  \frac{\lambda\,|\det B|\, \mathcal{C}_{\mathbf{k-m}}(\boldsymbol{\omega})}{(2\pi)^2} \int_{\mathbb{R}^2} f_{\mathbf {m}}\left(\lambda {\bf x} + \frac{\lambda}{2} \boldsymbol{\xi}\right) \overline{f_\mathbf{k}\left(\lambda {\bf x} - \frac{\lambda}{2} \boldsymbol{\xi}\right)}  \exp\left\{i\, \boldsymbol{\omega}^T B \boldsymbol{\xi} \right\} d\boldsymbol{\xi}.\notag
\end{align*}
Now, change the variables with $\boldsymbol{\zeta} = \lambda \boldsymbol{\xi}$, where $\boldsymbol{\zeta} = (\zeta_{1},\zeta_{2})^{T}$ so that $d\boldsymbol{\xi} = \frac{d\boldsymbol{\zeta}}{\lambda^2}$
\begin{align*}
\mathcal{W}_{D_\lambda f}^\Omega\left(\bf{x}, \boldsymbol{\omega}\right) &=  \frac{\lambda\,|\det B|\, \mathcal{C}_{\mathbf{k-m}}(\boldsymbol{\omega})}{(2\pi)^2} \int_{\mathbb{R}^2} f_{\mathbf {m}}\left(\lambda \bf{x} + \frac{\boldsymbol{\zeta}}{2}\right) \overline{f_\mathbf{k}\left(\lambda \bf{x} - \frac{\boldsymbol{\zeta}}{2}\right)} \exp\left\{i\, \boldsymbol{\omega}^T B \frac{\boldsymbol{\zeta}}{\lambda} \right\} \frac{d\boldsymbol{\zeta}}{\lambda^2} \\
&=  \frac{|\det B|\, \mathcal{C}_{\mathbf{k-m}}(\boldsymbol{\omega})}{\lambda\,(2\pi)^2} \int_{\mathbb{R}^2} f_{\mathbf {m}}\left(\lambda \bf{x} + \frac{\boldsymbol{\zeta}}{2}\right) \overline{f_\mathbf{k}\left(\lambda \bf{x} - \frac{\boldsymbol{\zeta}}{2}\right)} \exp\left\{i\, \frac{\boldsymbol{\omega}^T}{\lambda} B \boldsymbol{\zeta}\right\} d\boldsymbol{\zeta}\\
&= \frac{1}{\lambda} \, \mathcal{W}_f^\Omega\left(\lambda \bf{x}, \frac{\boldsymbol{\omega}}{\lambda}\right).
\end{align*}
Hence, the proof is established.
\end{proof}

\textbf{9. Convolution property:} For any two functions $f, g \in L^{2}(\mathbb{R}^{2})$, the 2D-NSQPWD corresponding to their convolution $f*g$ can be expressed as  
\begin{align}\label{S3E21}
\mathcal{W}_{f*g}^{\Omega}\left({\bf{x}},\boldsymbol{\omega}\right)
&= \frac{ (2\pi)^2\,\mathcal{C}_{\mathbf{m-k}}(\boldsymbol{\omega})}{|\det B|}\int_{\mathbb{R}^2}
\mathcal{W}_f^{\Omega}\left({{\bf{u}}},\boldsymbol{\omega}\right)
\mathcal{W}_g^{\Omega}\left({\bf{x}}-{\bf{u}},\boldsymbol{\omega}\right)d{\bf{u}},
\end{align}
Let ${\bf u}=(u_1,u_2)^T$, $\boldsymbol{\zeta} = (\zeta_{1},\zeta_{2})^{T}$, $\boldsymbol{\zeta}' = (\zeta'_{1},\zeta'_{2})^{T}$,  
$\boldsymbol{\nu} = (\nu_{1},\nu_{2})^{T}$, and $\boldsymbol{\nu}' = (\nu'_{1},\nu'_{2})^{T}$.  
The convolution of $f$ and $g$ is given by
\begin{align*}
(f*g)({\bf{x}})&=\int_{\mathbb{R}^2} f(\boldsymbol{\zeta})\, 
g( {\bf{x}}  - \boldsymbol{\zeta})\,d{\boldsymbol{\zeta}}
\end{align*}
\begin{proof}
Based on Definition~\ref{S2D1}, we obtain
\begin{align*}
\mathcal{W}^\Omega_{f*g}\left({\bf x},\boldsymbol{\omega}\right) 
&=  \frac{|\det B|\,\mathcal{C}_{\mathbf{k-m}}({\boldsymbol{\omega}}) }{(2 \pi)^2} 
\int_{\mathbb{R}^2} (f*g)_{{\mathbf {m}}}\left({\bf{x}} +\tfrac{1}{2} {\boldsymbol{\xi}} \right) 
\overline{(f*g)_{\mathbf{k}}\left( {\bf{x}} - \tfrac{1}{2} {\boldsymbol{\xi}} \right)}
\exp\left\{i\, {\boldsymbol{\omega}^T}B {\boldsymbol{\xi}}\right\} d\boldsymbol{\xi}\notag \\
&= \frac{|\det B|\,\mathcal{C}_{\mathbf{k-m}}({\boldsymbol{\omega}}) }{(2 \pi)^2} 
\int_{\mathbb{R}^2} 
\left( \int_{\mathbb{R}^2} f_{{\mathbf {m}}}\left(\boldsymbol{\zeta}\right)\, 
g_{{\mathbf {m}}}\left( {\bf{x}} + \tfrac{1}{2}\boldsymbol{\xi} - \boldsymbol{\zeta} \right) 
\, d\boldsymbol{\zeta} \right) \notag \\
&\quad \times
\left( \int_{\mathbb{R}^2} \overline{ f_{\mathbf{k}}\left(\boldsymbol{\zeta}^\prime\right)} \overline{ g_{\mathbf{k}}\left( {\bf{x}} - \tfrac{1}{2}\boldsymbol{\xi} - \boldsymbol{\zeta}^\prime \right)} d\boldsymbol{\zeta}^\prime \right)
\exp\left\{i\, {\boldsymbol{\omega}^T}B {\boldsymbol{\xi}}\right\}d\boldsymbol{\xi} \notag\\
\end{align*}
Substituting $\zeta = \mathbf{u} + \tfrac{1}{2}{\boldsymbol{\nu}}$ , $\zeta' = \mathbf{u} - \tfrac{1}{2}{\boldsymbol{\nu}}$,
the measure transforms as $d\zeta\,d\zeta' = d\mathbf{u}\,d\boldsymbol{\nu}$ with a Jacobian equal to 1
\begin{align*}
\mathcal{W}^\Omega_{f*g}\left({\bf x},\boldsymbol{\omega}\right) 
&= \frac{|\det B|\,\mathcal{C}_{\mathbf{k-m}}(\boldsymbol{\omega})}{(2\pi)^2}
\int_{\mathbb{R}^2\times\mathbb{R}^2\times\mathbb{R}^2}
f_{\mathbf m}\left(\mathbf{u}+ \tfrac{1}{2} {\boldsymbol{\nu}}\right)
\overline{f_{\mathbf k}\left(\mathbf{u}- \tfrac{1}{2} {\boldsymbol{\nu}}\right)}\,g_{\mathbf m}\left(\bf{x}-\mathbf{u}+ \tfrac{1}{2} {\boldsymbol{\xi}}- \tfrac{1}{2} {\boldsymbol{\nu}}\right) \notag\\
&\quad\times \,
\overline{g_{\mathbf k}\left(\bf{x}-\mathbf{u}- \tfrac{1}{2} {\boldsymbol{\xi}}+ \tfrac{1}{2} {\boldsymbol{\nu}}\right)}
\exp\left\{i\,\boldsymbol{\omega}^T B\boldsymbol{\xi}\right\}d\boldsymbol{\xi}
d\mathbf{u}\,d\boldsymbol{\nu}.
\end{align*}

Again, by changing variables $\boldsymbol{\xi}=\boldsymbol{\nu}+\boldsymbol{\nu}^\prime$ and $d\boldsymbol{\xi}=d\boldsymbol{\nu}^\prime$
\begin{align*}
\mathcal{W}^\Omega_{f*g}\left({\bf x},\boldsymbol{\omega}\right) &= \frac{|\det B|\,\mathcal{C}_{\mathbf{k-m}}(\boldsymbol{\omega})}{(2\pi)^2}
\int_{\mathbb{R}^2\times\mathbb{R}^2\times\mathbb{R}^2}
f_{\mathbf m}\left(\mathbf{u}+ \tfrac{1}{2} {\boldsymbol{\nu}}\right)
\overline{f_{\mathbf k}\left(\mathbf{u}- \tfrac{1}{2} {\boldsymbol{\nu}}\right)}\,g_{\mathbf m}\left(\bf{x}-\mathbf{u}+ \tfrac{1}{2} {\boldsymbol{\nu}^\prime}\right) \notag\\
&\quad\times \,
\overline{g_{\mathbf k}\left(\bf{x}-\mathbf{u}- \tfrac{1}{2} {\boldsymbol{\nu}^\prime}\right)}
\exp\left\{i\,\boldsymbol{\omega}^T B(\boldsymbol{\nu}+\boldsymbol{\nu}^\prime)\right\}d\boldsymbol{\nu}^\prime
d\mathbf{u}\,d\boldsymbol{\nu}\notag\\
&= \frac{|\det B|\,\mathcal{C}_{\mathbf{k-m}}(\boldsymbol{\omega})}{(2\pi)^2}
\int_{\mathbb{R}^2}
\left(\int_{\mathbb{R}^2}f_{\mathbf m}\left(\mathbf{u}+ \tfrac{1}{2} {\boldsymbol{\nu}}\right)
\overline{f_{\mathbf k}\left(\mathbf{u}- \tfrac{1}{2} {\boldsymbol{\nu}}\right)}\exp\left\{i\,\boldsymbol{\omega}^T B\boldsymbol{\nu}\right\}d\boldsymbol{\nu}\right) \notag\\
&\quad\times\left(\int_{\mathbb{R}^2} g_{\mathbf m}\left(\bf{x}-\mathbf{u}+ \tfrac{1}{2} {\boldsymbol{\nu}^\prime}\right)
\overline{g_{\mathbf k}\left(\bf{x}-\mathbf{u}- \tfrac{1}{2} {\boldsymbol{\nu}^\prime}\right)}
\exp\left\{i\,\boldsymbol{\omega}^T B\boldsymbol{\nu}^\prime\right\}d\boldsymbol{\nu}^\prime\right)
d\mathbf{u}\notag\\
&= \frac{ (2\pi)^2\,\mathcal{C}_{\mathbf{m-k}}(\boldsymbol{\omega})}{|\det B|}\int_{\mathbb{R}^2}
\mathcal{W}_f^{\Omega}\left({{\bf{u}}},\boldsymbol{\omega}\right)
\mathcal{W}_g^{\Omega}\left({\bf{x}}-{\bf{u}},\boldsymbol{\omega}\right)d{\bf{u}}.
\end{align*}
Hence, the proof is established.
\end{proof}

\section{Applications}\label{S4}

As an extended form of the WD, NSQPWD offers a wide range of applications in signal processing. One of its key advantages is the ability to handle 2D-LFM signals directly. To highlight the strength of the proposed theoretical framework, we apply the NSQPWD to the detection of 2D-LFM signals. In the simulation study, Two scenarios are analyzed, namely a mono-component 2D-LFM signal and multi-component cases involving bi-component and tri-component signals. Results from both settings clearly demonstrate the practicality, efficiency, and robustness of the new definition in real-world detection applications\cite{Lahiri,Zhang1}.

\subsection{2D-LFM signal with a single component:}Let a single-component LFM signal be defined as:
\begin{align}\label{S4E1}
	\mathcal{S}({\bf x}) = \kappa_0 \exp\big\{i \left(\left(\alpha_0 x_1 + \beta_0 x_1^2\right) + \left(\mu_0 x_2 + \lambda_0 x_2^2\right) \right)\big\},\quad {\bf x} \in \left[ -\frac{T}{2}, \frac{T}{2} \right] \times \left[ -\frac{T}{2}, \frac{T}{2} \right],
\end{align}
where  $\kappa_0$ denotes the amplitude, $\alpha_0$, $\mu_0$ represent the initial frequencies  $\beta_0\neq 0$, $\lambda_0\neq 0$ are the corresponding frequency modulation rates. It follows that
\begin{align}\label{S4E2}
	&\mathcal{S}_\mathbf{m}\left( {\bf{x}} + \tfrac{1}{2} {\boldsymbol{\xi}} \right) 
		\overline{\mathcal{S}_\mathbf{k}\left( {\bf{x}} - \tfrac{1}{2} {\boldsymbol{\xi}} \right)}\notag\\
	&= |\kappa_0|^2\, C_\mathbf{m-k}({\bf x})\, C_\mathbf{m-k}^\frac{1}{4}(\boldsymbol{\xi}) \exp\left\{i\,((\alpha_0 \,\tau +\mu_0\, \eta ) +2 (\beta_0\, x_1\, \tau +\lambda_0\,x_2\,\eta))\right\} \notag
	\\
	&\quad\times\exp\left\{i\left((m_1+k_1)x_1\tau+(m_2+k_2)\frac{x_1\eta+x_2\tau}{2}+(m_3+k_3)x_2 \eta+(m_4+3k_4)\frac{\tau}{4}\right.\right.\notag\\
	&\qquad \left. \left.+(m_5+3k_5){\frac{\eta}{4}}\right)\right\}.
\end{align}
Using \eqref{S4E2} and applying Definition \ref{S2D1}, we have
\begin{align*}
	&\mathcal{W}_\mathcal{S}^{\Omega}\left(\bf{x}, \boldsymbol{\omega}\right)\notag\\ 
	&=  \frac{|\det B|\,\mathcal{C}_{\mathbf{k-m}}(\boldsymbol{\omega}) }{(2 \pi)^2} \int_{\mathbb{R}^2} \mathcal{S}_\mathbf{m}\left( {\bf{x}} + \tfrac{1}{2} {\boldsymbol{\xi}} \right) 
	\overline{\mathcal{S}_\mathbf{k}\left( {\bf{x}} - \tfrac{1}{2} {\boldsymbol{\xi}} \right)}\,\exp\left\{i\, \boldsymbol{\omega}^T B \boldsymbol{\xi} \right\} \, d\boldsymbol{\xi}\notag
	\\
	&= \frac{|\kappa_0|^2\,|\det B|\,\mathcal{C}_\mathbf{m-k}({\bf{x}})\,\mathcal{C}_{\mathbf{k-m}}({\boldsymbol{\omega}}) }{(2 \pi)^2} \int_{-\frac{T}{2}}^{\frac{T}{2}} \int_{-\frac{T}{2}}^{\frac{T}{2}} C_\mathbf{m-k}^\frac{1}{4}(\boldsymbol{\xi})\notag\\
	&\quad \times \exp\left\{i \left(	 (m_1 + k_1) x_1 + \frac{(m_2 + k_2)}{2} x_2+{\frac{(m_4+3k_4)}{4}}+ \alpha_0 + 2\beta_0 x_1 + b_{11} \omega_1+ b_{12}\omega_2 \right) \tau \right\}\notag\\
	&\quad\times \exp\left\{i \left( \frac{(m_2 + k_2)}{2} x_1+ (m_3 + k_3)x_2 +{\frac{(m_5+3k_5)}{4}}+ \mu_0 + 2\lambda_0 x_2 +b_{12} \omega_1+ b_{22}\omega_2 \right) \eta \right\} \, d\tau d\eta.
\end{align*} 
When $\mathbf{m} = \mathbf{k}$, which means $m_1 = k_1,\, m_2 = k_2,\, m_3 = k_3,\, m_4 = k_4,\, m_5 = k_5$, we infer directly that
\begin{align}\label{S4E3}
	\mathcal{W}_\mathcal{S}^{\Omega}\left(\bf{x}, \boldsymbol{\omega}\right) 
	&=  \frac{|\kappa_0|^2\,|\det B| }{(2 \pi)^2} \int_{-\frac{T}{2}}^{\frac{T}{2}} \int_{-\frac{T}{2}}^{\frac{T}{2}}  \exp \left\{i\,( 2m_1 x_1 + m_2 x_2 + m_4+ \alpha_0 + 2\beta_0 x_1 + b_{11} \omega_1+ b_{12}\omega_2)\tau \right\} \notag \\
	&\quad \times \exp\left\{i\,( m_2  x_1 + 2m_3 x_2 +m_5+ \mu_0 + 2\lambda_0 x_2 +b_{12} \omega_1+ b_{22}\omega_2) \eta \right\} d\tau d\eta \notag
	\\
	&= \frac{|\kappa_0|^2\,|\det B| }{(2 \pi)^2} \int_{-\frac{T}{2}}^{\frac{T}{2}} \int_{-\frac{T}{2}}^{\frac{T}{2}}  \exp \left\{ i\,( (2m_1 + 2\beta_0)\, x_1 + m_2\, x_2 + b_{11} \omega_1 + b_{12} \omega_2 + m_4 + \alpha_0) \tau \right\} \notag\\
	&\quad \times \exp \left\{ i \,( m_2\, x_1 + (2m_3 + 2\lambda_0)\, x_2 + b_{12} \omega_1 + b_{22} \omega_2 + m_5 + \mu_0 ) \eta \right\} \, d\tau\, d\eta\notag
	\\
	&= \frac{|\kappa_0|^2\,|\det B| }{(2 \pi)^2} \left( \int_{-\frac{T}{2}}^{\frac{T}{2}} 
	\exp\left\{ i \left( (2m_1 + 2\beta_0) x_1 + m_2 x_2 + b_{11} \omega_1 + b_{12} \omega_2 + m_4 + \alpha_0 \right) \tau \right\} \, d\tau \right)\notag \\
	&\quad \times \left( \int_{-\frac{T}{2}}^{\frac{T}{2}} \exp\left\{ i \left(m_2 x_1 + (2m_3 + 2\lambda_0) x_2 + b_{12} \omega_1 + b_{22} \omega_2 + m_5 + \mu_0 \right) \eta \right\} \, d\eta \right)\notag
	\\
	&=\frac{|\kappa_0|^2\,T^2\,|\det B| }{(2 \pi)^2}\,  \text{sinc}\left\{ \frac{T}{2} \left( m_4 + m_2 x_2 + \alpha_0 + 2 x_1 (m_1 + \beta_0) + b_{11} \omega_1 + b_{12} \omega_2 \right)\right\}\notag\\
	&\quad \times \text{sinc}\left\{ \frac{T}{2} \left( m_5 + m_2 x_1 + \mu_0 + 2 x_2 (m_3 + \lambda_0) + b_{12} \omega_1 + b_{22} \omega_2 \right)\right\}.
\end{align} 
Equations \eqref{S4E3} demonstrate that, by selecting an appropriate value for the parameter $\Omega$, the 2D-NSQPWD can effectively reshape a single 2D-LFM component $\mathcal{S}(\bf x)$ into a localized pulse-like representation. Reveals potential of the 2D-NQPWD as a powerful tool for the analysis and detection of single-component 2D-LFM signals. To examine this behavior, we consider the following form of a single-component 2D-LFM signal:
\begin{align*}
	\mathcal{S}(x_1,x_2) =  \exp \left\{i\, ( (0.3 x_1 + 0.2 x_1^2) + (0.1 x_2 + 0.5x_2^2) )\right\} , \qquad T=40
\end{align*}
and the parameter 
\begin{align*}
	\Omega_0 = \Bigg\{ & A = 
	\begin{bmatrix}
		1 & -5 \\
		5 & 1
	\end{bmatrix}, \,
	B = 
	\begin{bmatrix}
		2 & 1 \\
		1 & 4
	\end{bmatrix},  \,
	 C = 
	\begin{bmatrix}
		1 & \frac{-13}{7} \\
		\frac{13}{7} & 1
	\end{bmatrix}, \,
	D = 
	\begin{bmatrix}
		2 & 1 \\
		2 & 5
	\end{bmatrix}, \,
	E = 
	\begin{bmatrix}
		1 & 2 \\
		3 & 4
	\end{bmatrix}
	\Bigg\}.
\end{align*}

\parindent=0mm\vspace{.1in}
The 2D-NSQPWD of $\mathcal{S}(\bf x)$ at specific values of $ \bf x$ is presented in Fig.~\ref{F1}. Figures 1(a), 1(c), and 1(e) depict 3D surface plots representing the distribution of signal energy in the time-frequency plane, characterized by sharp peaks and detailed features indicative of complex frequency components. Figures 1(b), 1(d), and 1(f) show corresponding 2D contour plots that delineate high-energy regions and ridges, facilitating the visualization of instantaneous frequency trajectories and signal structure. The combination of surface and contour representations highlights the improved time-frequency concentration and resolution provided by the 2D-NSQPWD within the QPFT domain.
\begin{figure}[htbp]
	\centering		
	\begin{minipage}[b]{0.49\textwidth}
		\centering
		\includegraphics[width=\textwidth]{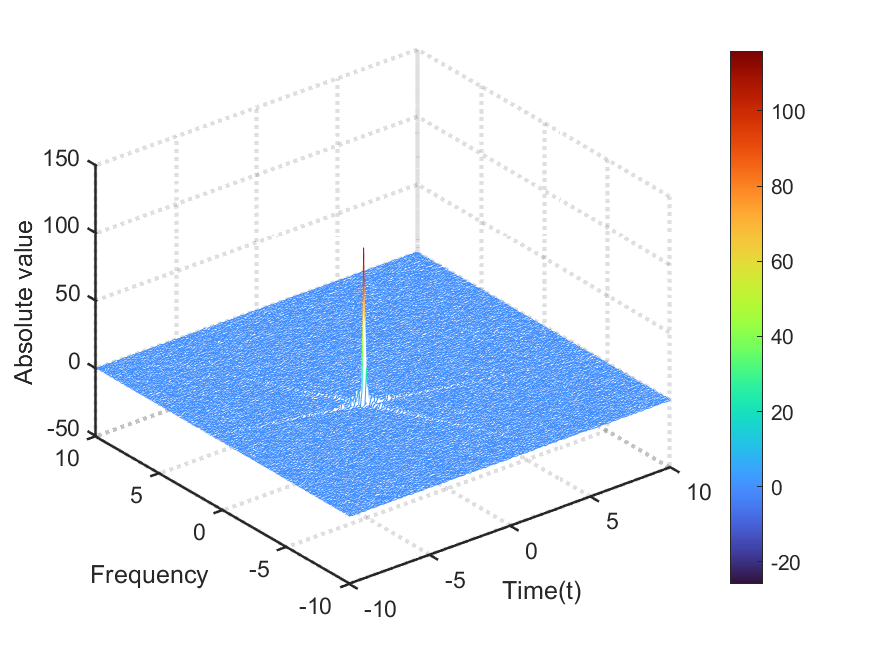}
		{\rm (a) $	\mathcal{W}_\mathcal{S}^{\Omega_0}(0.40,0.10, \omega_1,\omega_2)$.}
	\end{minipage}
	\hfill
	\begin{minipage}[b]{0.49\textwidth}
		\centering
		\includegraphics[width=\textwidth]{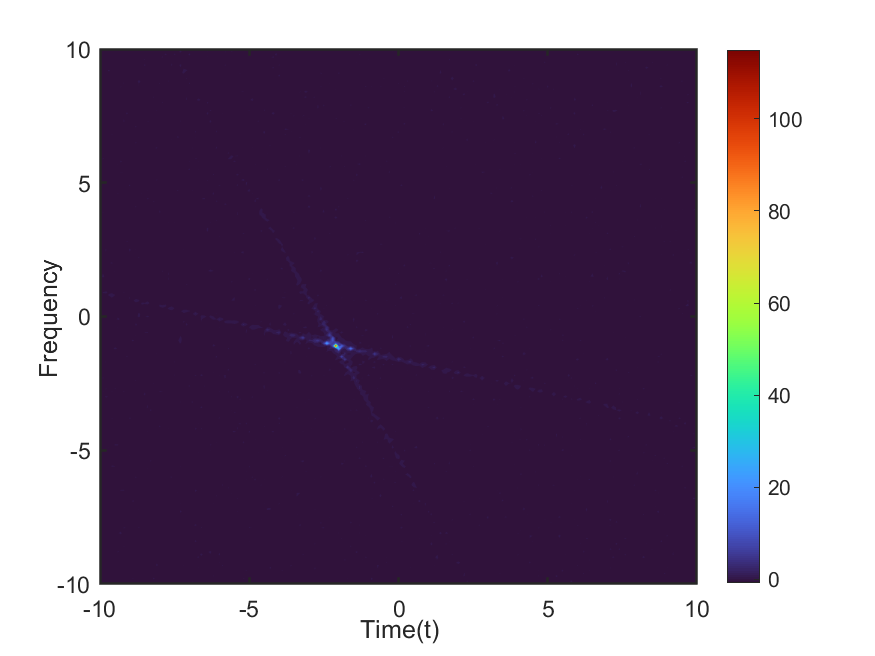}
		{\rm (b)  $\mathcal{W}_\mathcal{S}^{\Omega_0}(0.40,0.10, \omega_1,\omega_2)$ . }
	\end{minipage}
\end{figure}	 
\begin{figure}[htbp]
	\centering		
	\begin{minipage}[b]{0.49\textwidth}
		\centering
		\includegraphics[width=\textwidth]{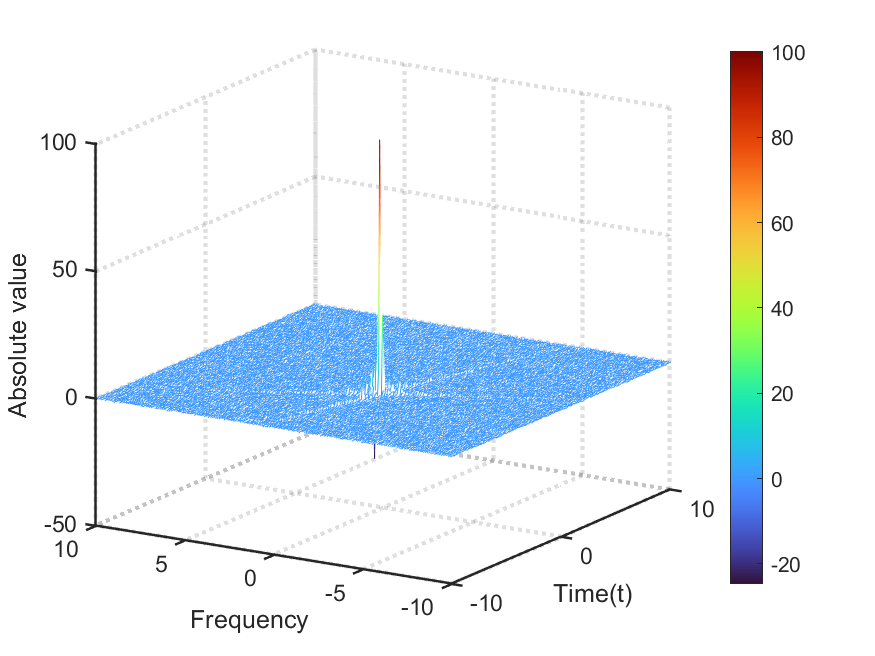}
		{\rm (c) $\mathcal{W}_\mathcal{S}^{\Omega_0}(0.60,0.30, \omega_1,\omega_2)$.}
	\end{minipage}
	\hfill
	\begin{minipage}[b]{0.49\textwidth}
		\centering
		\includegraphics[width=\textwidth]{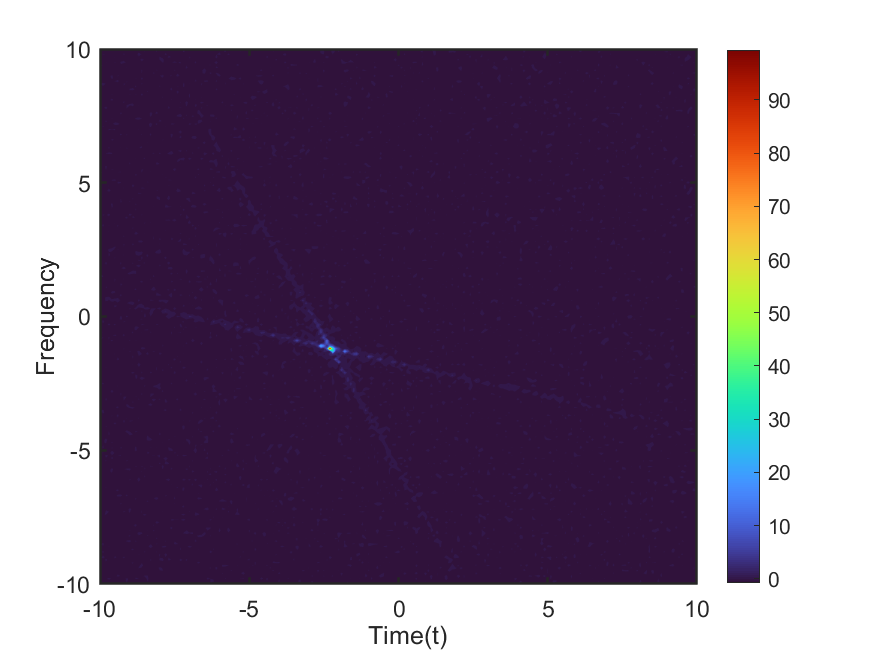}
		{\rm (d) $\mathcal{W}_\mathcal{S}^{\Omega_0}(0.60,0.30, \omega_1,\omega_2)$ .}
	\end{minipage}
\end{figure}	 	 
\begin{figure}[htbp]
	\centering		
	\begin{minipage}[b]{0.49\textwidth}
		\centering
		\includegraphics[width=\textwidth]{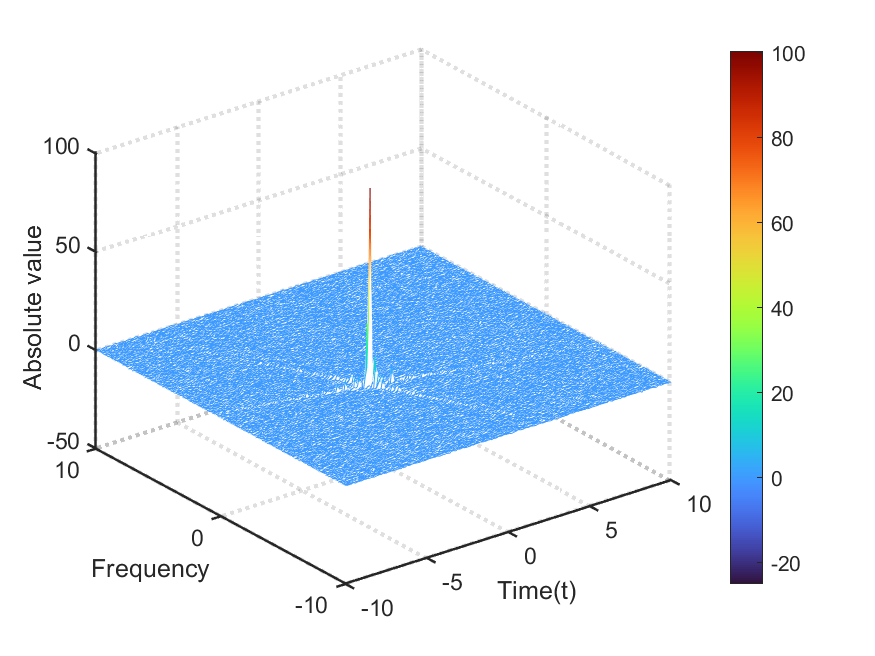}
		{\rm (e) $\mathcal{W}_\mathcal{S}^{\Omega_0}(0.40,0.50, \omega_1,\omega_2)$.}
	\end{minipage}
	\hfill
	\begin{minipage}[b]{0.49\textwidth}
		\centering
		\includegraphics[width=\textwidth]{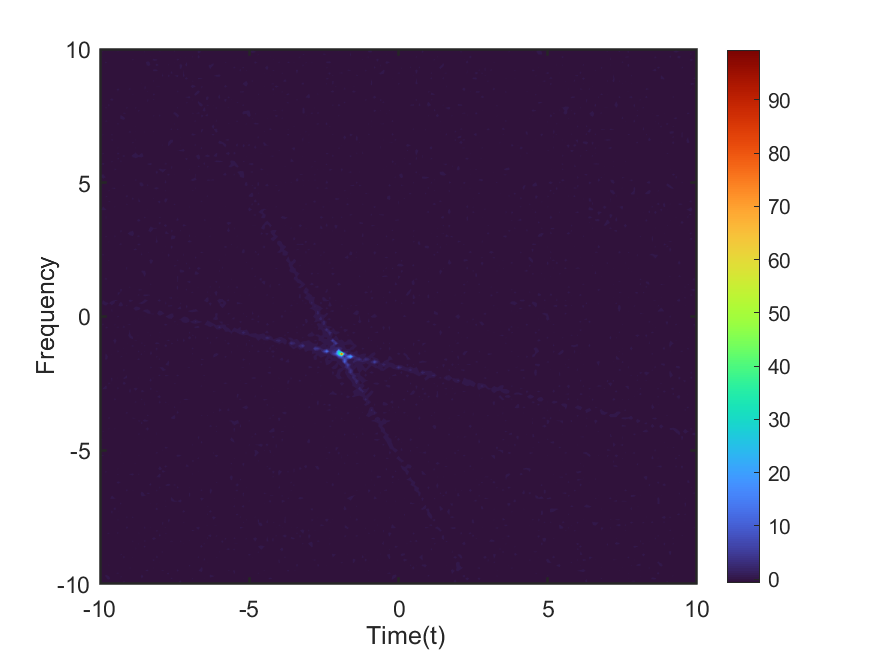}
		{\rm (f)   $\mathcal{W}_\mathcal{S}^{\Omega_0}(0.40,0.50, \omega_1,\omega_2)$.}
	\end{minipage}
	\caption{\small The 2D-NSQPWD and corresponding contour plots of the signal $\mathcal{S}(\bf x)$ presented at an SNR of 10 dB for various values of $ \bf x $ and matrix parameters $\Omega_0$. }
	\label{F1}
\end{figure}	 	 

\subsection{Bi-component LFM signal:}
Consider a bi-component LFM signal defined as
\begin{align}\label{S4E4}
	\mathcal{U}({\bf x}) =  \mathcal{U}_1({\bf x}) + \mathcal{U}_2({\bf x}), \quad {\bf x} \in \left[ -\frac{T}{2}, \frac{T}{2} \right] \times \left[ -\frac{T}{2}, \frac{T}{2} \right],
\end{align}
where $\mathcal{U}_1({\bf x}) = \kappa_1 \exp\left\{i\, ( (\alpha_1 x_1 + \beta_1 x_1^{2}) + (\mu_1 x_2 + \lambda_1 x_2^{2})) \right\}$  and \\  $\mathcal{U}_2({\bf x})=\kappa_2 \exp\left\{i\,( (\alpha_2 x_1 + \beta_2 x_1^{2}) + (\mu_2 x_2 + \lambda_2 x_2^{2}) )\right\}$. The 2D-NSQPWD of  $\mathcal{U}({\bf x})$ is given by
\begin{align}\label{S4E5}
	\mathcal{W}_\mathcal{U}^{\Omega}\left(\bf{x}, \boldsymbol{\omega}\right)  =  \mathcal{W}^{\Omega}_{ \mathcal{U}_{1}}\left(\bf{x}, \boldsymbol{\omega}\right)+  \mathcal{W}^{\Omega}_{\mathcal{U}_{2}}\left(\bf{x}, \boldsymbol{\omega}\right) +\mathcal{W}^{\Omega}_{\mathcal{U}_{1}, \mathcal{U}_{2}}\left(\bf{x}, \boldsymbol{\omega}\right)+ \mathcal{W}^{\Omega}_{\mathcal{U}_{2}, \mathcal{U}_{1}}\left(\bf{x}, \boldsymbol{\omega}\right).
\end{align}	
In this expression, the first two terms represent the auto-terms corresponding to the individual components $\mathcal{U}_1({\bf x})$ and $\mathcal{U}_2({\bf x})$, while the remaining two terms are cross-terms. The NQPWD of these cross-terms can be expressed as:
\begin{align}\label{S4E6}
	\mathcal{W}^{\Omega}_{\mathcal{U}_{1}, \mathcal{U}_{2}}\left(\bf{x}, \boldsymbol{\omega}\right)&=\frac{|\det B|\,\mathcal{C}_{\mathbf{k-m}}(\omega) }{(2 \pi)^2} \int_{\mathbb{R}^2} \mathcal{U}_{1,m}\left( {\bf{x}} + \tfrac{1}{2} {\boldsymbol{\xi}} \right) 
	\overline{\mathcal{U}_{2,k}\left( {\bf{x}} - \tfrac{1}{2} {\boldsymbol{\xi}} \right)}\,\exp\left\{i\, \boldsymbol{\omega}^T B \boldsymbol{\xi} \right\} \, d\boldsymbol{\xi}.
\end{align}
It is observed that
\begin{align}\label{S4E7}
	&\mathcal{U}_{1,m}\left({\bf{x}} + \tfrac{1}{2} {\boldsymbol{\xi}}  \right) 	\overline{\mathcal{U}_{2,k}\left({\bf{x}} - \tfrac{1}{2} {\boldsymbol{\xi}}  \right) }\notag\\
	&= \kappa_1\overline{\kappa_2}\, C_\mathbf{m-k}({\bf x)}\, C_\mathbf{m-k}^\frac{1}{4}({\boldsymbol{\xi}})\notag \\
	&\quad \times \exp\left\{i \left((\beta_1-\beta_2) x_1^2 +(\lambda_1-\lambda_2) x_2^2 +(\alpha_1-\alpha_2) x_1 +(\mu_1-\mu_2) x_2 \right)\right\}  \notag \\
	&\quad \times \exp\left\{i \left( \frac{(\beta_1-\beta_2)}{4} \tau^2 + \frac{(\lambda_1-\lambda_2)}{4} \eta^2 \right)\right\}\exp\left\{i \left( (m_1 + k_1) x_1  +{\frac{(m_4+3k_4)}{4}}\right.\right.\notag\\
	&\quad\left.\left.+ \frac{(m_2 + k_2)}{2} x_2+ \frac{(\alpha_1+\alpha_2)}{2} + (\beta_1+\beta_2) x_1 \right) \tau \right\} \exp\left\{i \left( \frac{(m_2 + k_2)}{2} x_1 \right.\right.\notag\\
	&\quad\left.\left.+ (m_3 + k_3)x_2 +{\frac{(m_5+3k_5)}{4}}+ \frac{(\mu_1+\mu_2)}{2} + (\lambda_1+\lambda_2) x_2 \right) \eta \right\}.
\end{align}
From equations \eqref{S4E6} and \eqref{S4E7} we get 
\begin{align*}
	&\mathcal{W}^{\Omega}_{\mathcal{U}_{1}, \mathcal{U}_{2}}\left(\bf{x}, \boldsymbol{\omega}\right)\notag\\
	&= \frac{ \kappa_1\overline{\kappa_2}\,|\det B|\,C_\mathbf{k-m}(\omega_1,\omega_2) C_\mathbf{m-k}(x_1,x_2)}{4\pi^2}\, 
	\exp\left\{i\,((\beta_1-\beta_2) x_1^2 +(\lambda_1-\lambda_2) x_2^2 +(\alpha_1-\alpha_2) x_1 \right.\notag\\
	& \quad \left.+(\mu_1-\mu_2) x_2 )\right\} \int_{-\frac{T}{2}}^{\frac{T}{2}} \int_{-\frac{T}{2}}^{\frac{T}{2}} C_\mathbf{m-k}^{\frac{1}{4}} (\tau,\eta)\,
	\exp\left\{i \left( \frac{(\beta_1-\beta_2)}{4} \tau^2 + \frac{(\lambda_1-\lambda_2)}{4} \eta^2 \right)\right\} \notag \\
	&\quad \times 	\exp\left\{i \left( (m_1 + k_1) x_1 + \frac{(m_2 + k_2)}{2} x_2 +{\frac{(m_4+3k_4)}{4}}+ \frac{(\alpha_1+\alpha_2)}{2} + (\beta_1+\beta_2) x_1 \right.\right.\notag \\
	&\quad \left.\left. + b_{11} \omega_1+b_{12}\omega_2 \right) \tau \right\} \exp\left\{i \left( \frac{(m_2 + k_2)}{2} x_1 + (m_3 + k_3)x_2 +{\frac{(m_5+3k_5)}{4}}+ \frac{(\mu_1+\mu_2)}{2} \right.\right.\notag \\
	&\quad \left.\left.+ (\lambda_1+\lambda_2) x_2 +b_{12} \omega_1+ b_{22}\omega_2 \right) \eta \right\}d\tau d\eta.
\end{align*} 
When $\mathbf{m} = \mathbf{k}$, $ \beta_1=\beta_2$, $\lambda_1=\lambda_2 $, we obtain 
\begin{align}\label{S4E8}
	&\mathcal{W}^{\Omega}_{\mathcal{U}_{1}, \mathcal{U}_{2}}\left(\bf{x}, \boldsymbol{\omega}\right)\notag\\
	&=  \frac{\kappa_1\overline{\kappa_2}\,|\det B|}{(2 \pi)^2}\,\exp\left\{i \left( (\alpha_1-\alpha_2) x_1 +(\mu_1-\mu_2) x_2 \right)\right\}\int_{-\frac{T}{2}}^{\frac{T}{2}} \int_{-\frac{T}{2}}^{\frac{T}{2}} \exp\left\{i \left( 2m_1 x_1 + m_2 x_2 + m_4\right.\right.\notag \\
    &\quad\left.\left. +\frac{(\alpha_1+\alpha_2)}{2} + 2\beta_1 x_1 + b_{11} \omega_1+ b_{12}\omega_2 \right) \tau \right\}\exp\left\{i \left( m_2  x_1 + 2m_3 x_2 +m_5+ \frac{(\mu_1+\mu_2)}{2}\right.\right.\notag \\
    	&\quad \left.\left. + 2\lambda_1 x_2 +b_{12} \omega_1+ b_{22}\omega_2 \right) \eta \right\} \, d\tau d\eta \notag
	\\
	&=\frac{\kappa_1\overline{\kappa_2}\,T^2\,|\det B| }{(2 \pi)^2}\,\,e^{i \left( (\alpha_1-\alpha_2) x_1 +(\mu_1-\mu_2) x_2 \right)} \notag \\
	&\quad \times\text{sinc}\left\{ \frac{T}{2} \left( m_4 + m_2 x_2 +\frac{(\alpha_1+\alpha_2)}{2}  + 2 x_1 (m_1 + \beta_1) + b_{11} \omega_1 + b_{12} \omega_2 \right)\right\}\notag\\
	&\quad \times \text{sinc}\left\{ \frac{T}{2} \left( m_5 + m_2 x_1 + \frac{(\mu_1+\mu_2)}{2}  + 2 x_2 (m_3 + \lambda_1) + b_{12} \omega_1 + b_{22} \omega_2 \right)\right\}
\end{align} 
Similarly, we obtain the following:
\begin{align}\label{S4E9}
	\mathcal{W}^{\Omega}_{\mathcal{U}_{2}, \mathcal{U}_{1}}\left(\bf{x}, \boldsymbol{\omega}\right)
	&=\frac{\kappa_2\overline{\kappa_1}\,T^2\,|\det B| }{(2 \pi)^2}\,\,e^{i \left( (\alpha_2-\alpha_1) x_1 +(\mu_2-\mu_1) x_2 \right)} \notag \\
	&\quad \times\text{sinc}\left\{ \frac{T}{2} \left( m_4 + m_2 x_2 +\frac{(\alpha_1+\alpha_2)}{2}  + 2 x_1 (m_1 + \beta_1) + b_{11} \omega_1 + b_{12} \omega_2 \right)\right\}\notag\\
	&\quad \times \text{sinc}\left\{ \frac{T}{2} \left( m_5 + m_2 x_1 + \frac{(\mu_1+\mu_2)}{2}  + 2 x_2 (m_3 + \lambda_1) + b_{12} \omega_1 + b_{22} \omega_2 \right)\right\}
\end{align} 
Therefore, the NSQPWD of the bi-component signal ${\mathcal{U}_{1}{(\bf{x})}+\mathcal{U}_{2}{(\bf{x})}}$, equation \eqref{S4E3},\eqref{S4E8} and \eqref{S4E9} put into \eqref{S4E5} as follows:
\begin{align}\label{S4E10}
	&\mathcal{W}^{\Omega}_{\mathcal{U}}\left(\bf{x}, \boldsymbol{\omega}\right)\notag\\
	&=\frac{|\kappa_1|^2\,T^2\,|\det B| }{(2 \pi)^2}\,  \text{sinc}\left\{ \frac{T}{2} \left( m_4 + m_2 x_2 + \alpha_1 + 2 x_1 (m_1 + \beta_1) + b_{11} \omega_1 + b_{12} \omega_2 \right)\right\}\notag\\
	&\quad \times \text{sinc}\left\{ \frac{T}{2} \left( m_5 + m_2 x_1 + \mu_1 + 2 x_2 (m_3 + \lambda_1) + b_{12} \omega_1 + b_{22} \omega_2 \right)\right\}\notag\\
	&\quad+\frac{|\kappa_2|^2\,T^2\,|\det B| }{(2 \pi)^2}\,  \text{sinc}\left\{ \frac{T}{2} \left( m_4 + m_2 x_2 + \alpha_2 + 2 x_1 (m_1 + \beta_2) + b_{11} \omega_1 + b_{12} \omega_2 \right)\right\}\notag\\
	&\quad \times \text{sinc}\left\{ \frac{T}{2} \left( m_5 + m_2 x_1 + \mu_2 + 2 x_2 (m_3 + \lambda_2) + b_{12} \omega_1 + b_{22} \omega_2 \right)\right\}\notag\\
	&\quad+\frac{\kappa_1\overline{\kappa_2}\,T^2\,|\det B| }{(2 \pi)^2}\,\,e^{i \left( (\alpha_1-\alpha_2) x_1 +(\mu_1-\mu_2) x_2 \right)} \notag \\
	&\quad \times\text{sinc}\left\{ \frac{T}{2} \left( m_4 + m_2 x_2 +\frac{(\alpha_1+\alpha_2)}{2}  + 2 x_1 (m_1 + \beta_1) + b_{11} \omega_1 + b_{12} \omega_2 \right)\right\}\notag\\
	&\quad \times \text{sinc}\left\{ \frac{T}{2} \left( m_5 + m_2 x_1 + \frac{(\mu_1+\mu_2)}{2}  + 2 x_2 (m_3 + \lambda_1) + b_{12} \omega_1 + b_{22} \omega_2 \right)\right\}\notag\\
	&\quad+\frac{\kappa_2\overline{\kappa_1}\,T^2\,|\det B| }{(2 \pi)^2}\,\,e^{i \left( (\alpha_2-\alpha_1) x_1 +(\mu_2-\mu_1) x_2 \right)} \notag \\
	&\quad \times\text{sinc}\left\{ \frac{T}{2} \left( m_4 + m_2 x_2 +\frac{(\alpha_1+\alpha_2)}{2}  + 2 x_1 (m_1 + \beta_1) + b_{11} \omega_1 + b_{12} \omega_2 \right)\right\}\notag\\
	&\quad \times \text{sinc}\left\{ \frac{T}{2} \left( m_5 + m_2 x_1 + \frac{(\mu_1+\mu_2)}{2}  + 2 x_2 (m_3 + \lambda_1) + b_{12} \omega_1 + b_{22} \omega_2 \right)\right\}.
\end{align}
As shown in Equation \eqref{S4E10}, the 2D-NSQPWD technique effectively identifies bi-component 2D-LFM signals, enabling precise detection and analysis. The signal under consideration is defined as:
\begin{align}
	\mathcal{U}({\bf x})&=4\,e^{i \left(0.2x_1 + 0.05x_1^2 + 0.15x_2 + 0.04x_2^2\right)} + 
	e^{i \left(0.4x_1 + 0.05 x_1^2 + 0.2x_2 + 0.04x_2^2\right)},\qquad T=40
\end{align}
with the matrix parameter $ \Omega_0$ and some special value of ${\bf x}$.
Fig.~\ref{F2} presents both 3D surface plots and 2D contour maps that visualize the real and imaginary components of the complex function $\mathcal{U}({\bf x})$. These visualizations reveal intricate oscillatory patterns and phase behaviors inherent to the bi-component LFM signal, offering insight into the signal’s structure across time and frequency dimensions. Fig.~\ref{F3} displays each scenario with a 3D magnitude plot on the left and its corresponding 2D contour on the right. The bi-component structure is evident as two distinct diagonal auto-terms in the contour plots, representing the instantaneous frequency trajectories of the chirps. At $ {\bf{x}} = (0.60, 0.10)$ , Figures 3(a) and 3(b) show that the energy is sharply localized, yielding clear auto-terms with minimal cross-term interference. For $ {\bf{x}} = (0.40, 0.30)$, Figures 3(c) and 3(d) show that the energy remains concentrated but shows slight frequency-axis broadening along with the emergence of faint cross-term artifacts. At $ {\bf{x}} = (0.20, 0.30)$, Figures 3(e) and 3(f) show that energy concentration decreases moderately, with more pronounced cross-term artifacts visible.

\begin{figure}[htbp]
	\centering		
	\begin{minipage}[b]{0.49\textwidth}
		\centering
		\includegraphics[width=\textwidth]{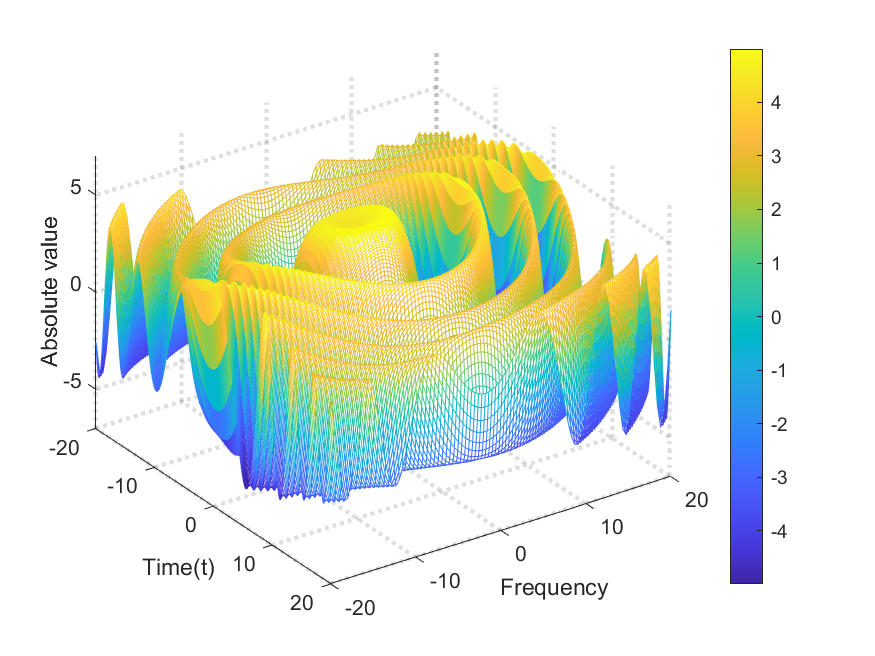}
		{\rm (a) Real part of $\mathcal{U}({\bf x})$.}
	\end{minipage}
	\hfill
	\begin{minipage}[b]{0.49\textwidth}
		\centering
		\includegraphics[width=\textwidth]{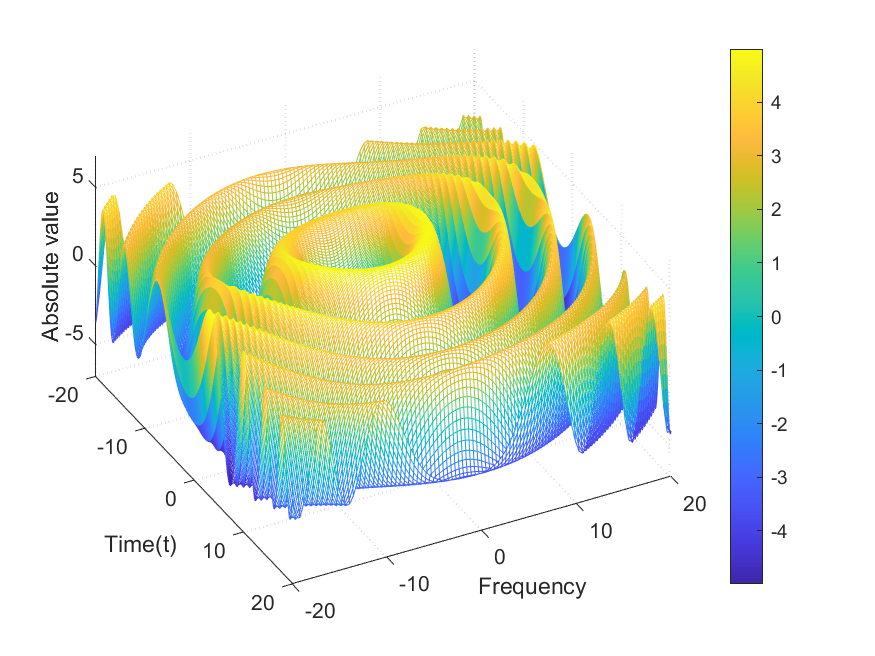}
		{\rm (b) Imaginary part of $\mathcal{U}({\bf x})$. }
	\end{minipage}
\end{figure}
\begin{figure}[htbp]
	\centering		
	\begin{minipage}[b]{0.49\textwidth}
		\centering
		\includegraphics[width=\textwidth]{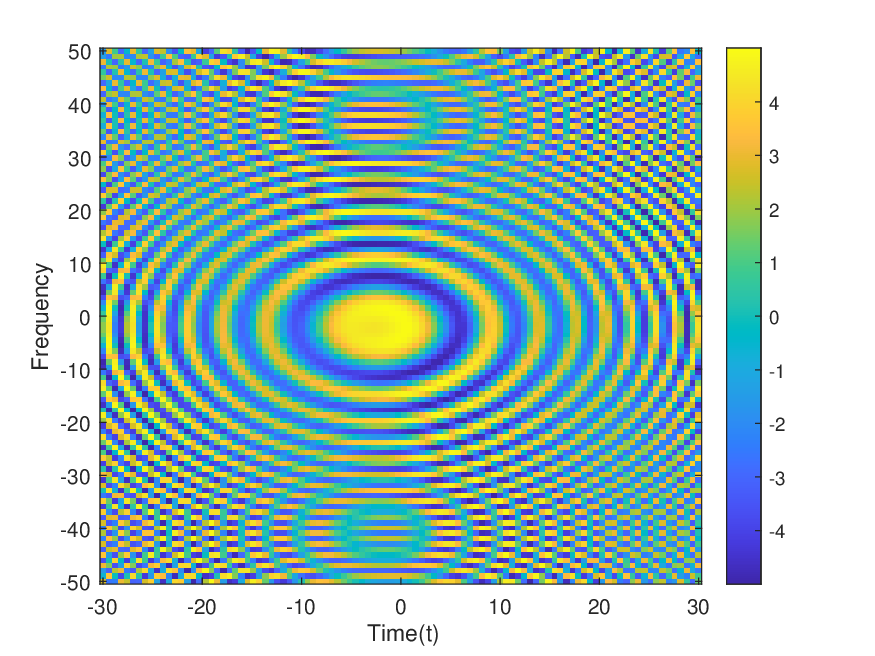}
		{\rm (c) 2D plot of the real part of $\mathcal{U}({\bf x})$.}
	\end{minipage}
	\hfill
	\begin{minipage}[b]{0.49\textwidth}
		\centering
		\includegraphics[width=\textwidth]{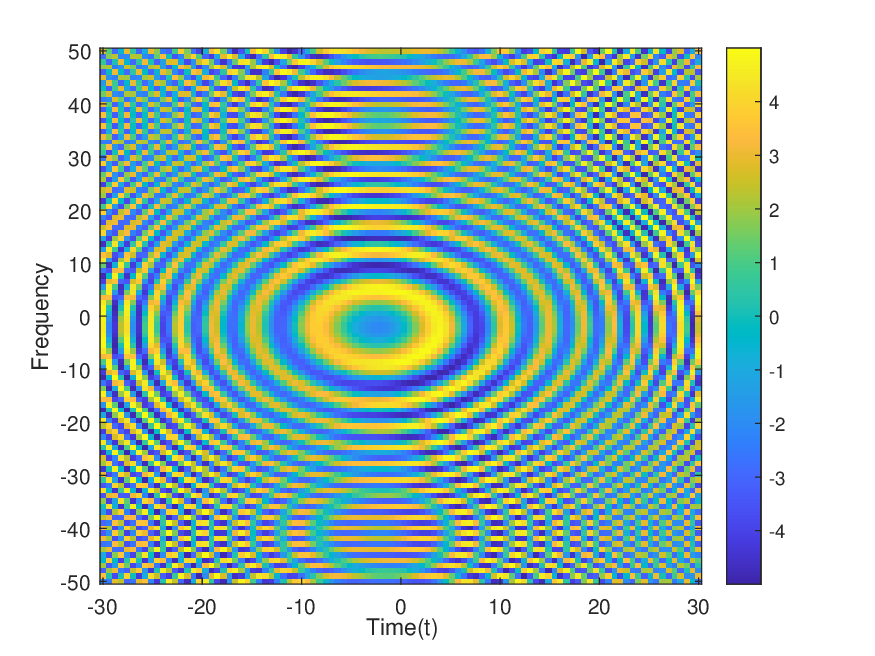}
		{\rm (d) 2D plot of the imaginary part of $\mathcal{U}({\bf x})$.}
	\end{minipage}
	\caption{\small Visualization of the complex function $\mathcal{U}(\bf{x})$ .}
	\label{F2}
\end{figure}

\begin{figure}[htbp]
	\centering		
	\begin{minipage}[b]{0.49\textwidth}
		\centering
		\includegraphics[width=\textwidth]{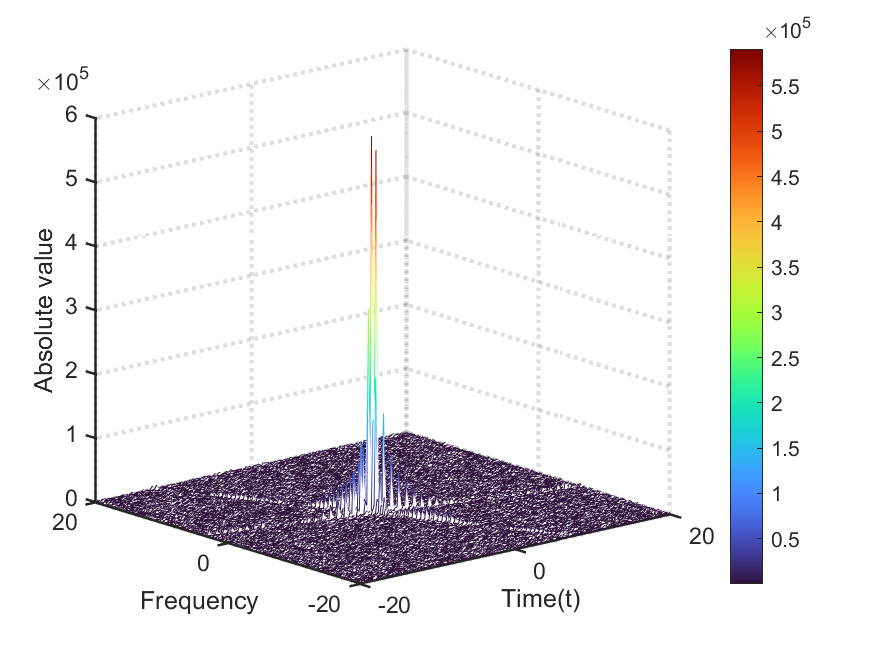}
		{\rm (a) $	\mathcal{W}_\mathcal{U}^{\Omega_0}(0.60,0.10, \omega_1,\omega_2)$.}
	\end{minipage}
	\hfill
	\begin{minipage}[b]{0.49\textwidth}
		\centering
		\includegraphics[width=\textwidth]{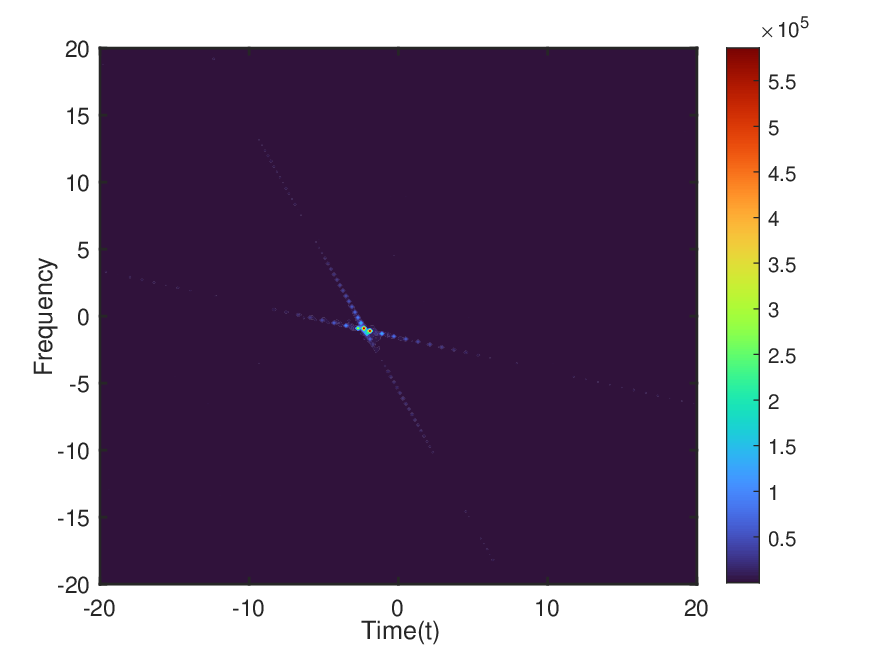}
		{\rm (b)  $\mathcal{W}_\mathcal{U}^{\Omega_0}(0.60,0.10, \omega_1,\omega_2)$ . }
	\end{minipage}
\end{figure}	 
\begin{figure}[htbp]
	\centering		
	\begin{minipage}[b]{0.49\textwidth}
		\centering
		\includegraphics[width=\textwidth]{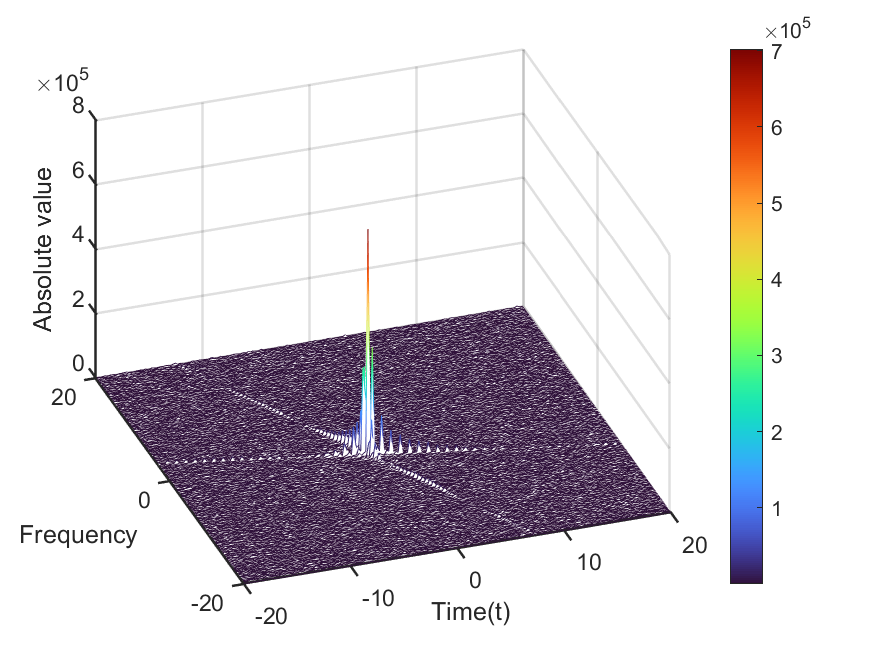}
		{\rm (c) $\mathcal{W}_\mathcal{U}^{\Omega_0}(0.40,0.30, \omega_1,\omega_2)$.}
	\end{minipage}
	\hfill
	\begin{minipage}[b]{0.49\textwidth}
		\centering
		\includegraphics[width=\textwidth]{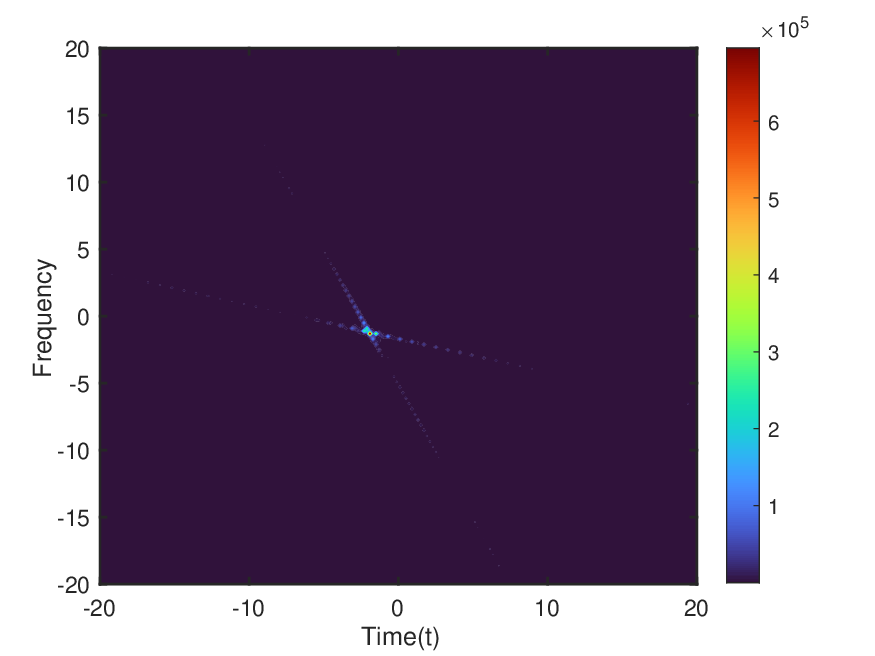}
		{\rm (d) $\mathcal{W}_\mathcal{U}^{\Omega_0}(0.40,0.30, \omega_1,\omega_2)$ .}
	\end{minipage}
\end{figure}	 	 
\begin{figure}[htbp]
	\centering		
	\begin{minipage}[b]{0.49\textwidth}
		\centering
		\includegraphics[width=\textwidth]{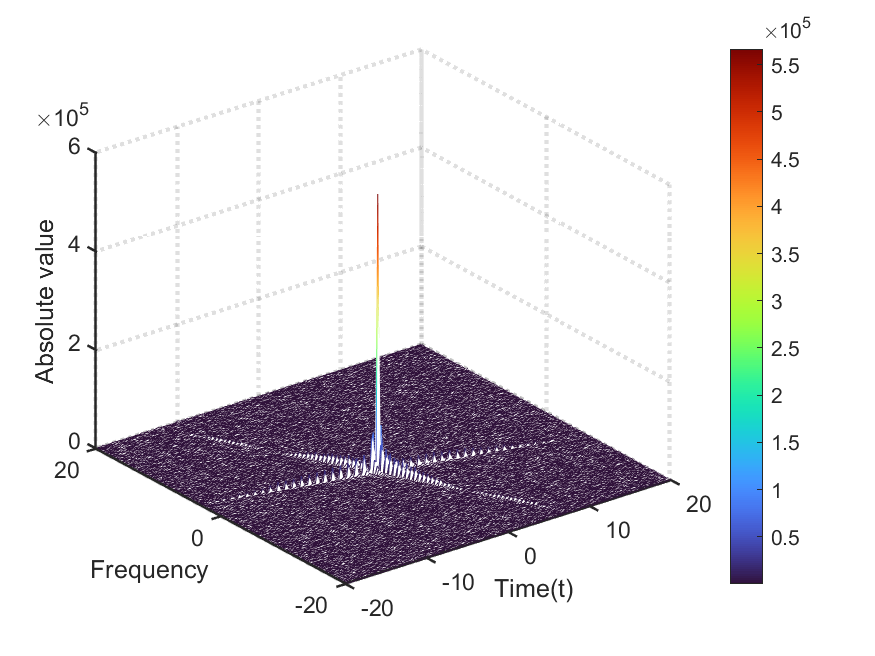}
		{\rm (e) $\mathcal{W}_\mathcal{U}^{\Omega_0}(0.20,0.30, \omega_1,\omega_2)$.}
	\end{minipage}
	\hfill
	\begin{minipage}[b]{0.49\textwidth}
		\centering
		\includegraphics[width=\textwidth]{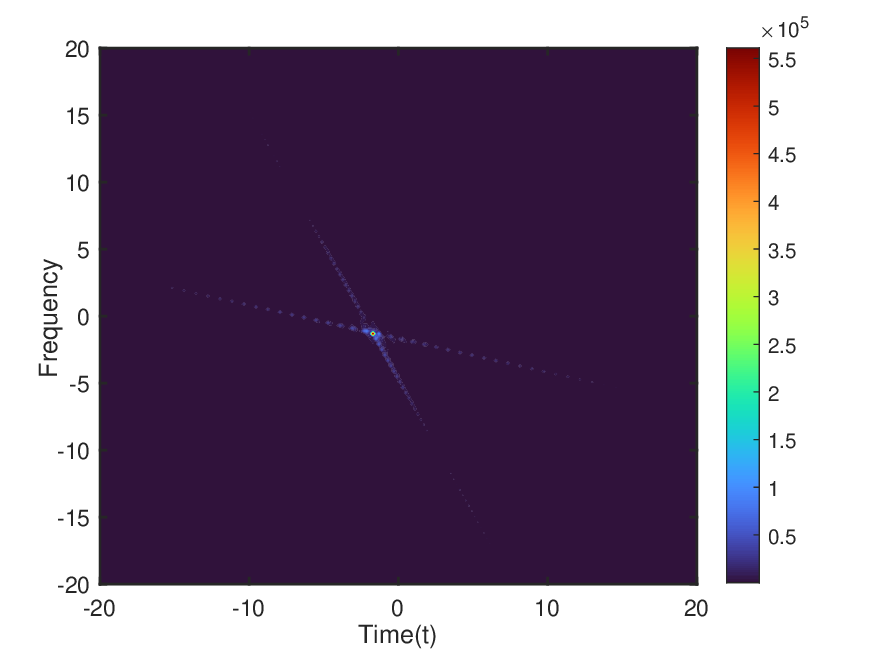}
		{\rm (f) $\mathcal{W}_\mathcal{U}^{\Omega_0}(0.20,0.30, \omega_1,\omega_2)$.}
	\end{minipage}
	\caption{\small The 2D-NSQPWD and corresponding contour plots of the bi-component signal $\mathcal{U}({\bf x})$ presented at an SNR of 10 dB for various values of $ \bf x $ and matrix parameters $\Omega_0$. }
	\label{F3}
\end{figure}

\subsection{Tri-component LFM signal:}
Consider a general tri-component LFM signal defined as
\begin{align}\label{S4E12}
    \mathcal{U}(\bf{x}) 
    &= \sum_{n=1}^{3} \mathcal{U}_n(\bf{x}), 
    \quad 
    \bf{x} \in \left[ -\frac{T}{2}, \frac{T}{2} \right]^2, \\
    \mathcal{U}_n(\bf{x}) 
    &= \kappa_n \exp\left\{ i\left[ (\alpha_n x_1 + \beta_n x_1^2) + (\mu_n x_2 + \lambda_n x_2^2) \right] \right\}, 
    \quad n = 1,\dots,3. \notag
\end{align}
 The 2D–NSQPWD corresponding to $\mathcal{U}(\bf{x})$ is given by
\begin{align}\label{S4E13}
    \mathcal{W}_{\mathcal{U}}^{\Omega}(\bf{x}, \boldsymbol{\omega}) 
    &= \sum_{n=1}^{3} \mathcal{W}^{\Omega}_{\mathcal{U}_n}(\bf{x}, \boldsymbol{\omega})
    + \sum_{\substack{n,m=1 \\ n\neq m}}^{3} 
      \mathcal{W}^{\Omega}_{\mathcal{U}_n,\,\mathcal{U}_m}(\bf{x}, \boldsymbol{\omega}).
\end{align}
In \eqref{S4E13}, the first summation corresponds to the auto-terms, describing the self time–frequency distribution of each individual component. For a single component $\mathcal{U}_n(\bf{x})$, the 2D–NSQPWD reduces to the form in \eqref{S4E3}. The second summation represents the cross-terms, capturing the time–frequency interaction between different components, evaluated using \eqref{S4E8}. Both contributions are computed through the 2D–NSQPWD framework in \eqref{S2E7}.

\parindent=0mm\vspace{.1in}
The operational sequence of the proposed 2D–NSQPWD for tri–component 2D–LFM signal detection is illustrated in Fig.~\ref{fig:2DNSQPWD_flow}. The diagram captures the logical progression from parameter initialization through signal synthesis and impulse response computation to the final detection output, providing a clear visual summary of the algorithm’s structure.

\parindent=0mm\vspace{.1in}
\tikzstyle{startstop} = [rectangle, rounded corners, minimum width=6cm, minimum height=1cm,text centered, draw=black, fill=blue!10]
\tikzstyle{process} = [rectangle, minimum width=6cm, minimum height=1cm, text centered, draw=black, fill=green!10]
\tikzstyle{arrow} = [thick,->,>=stealth]
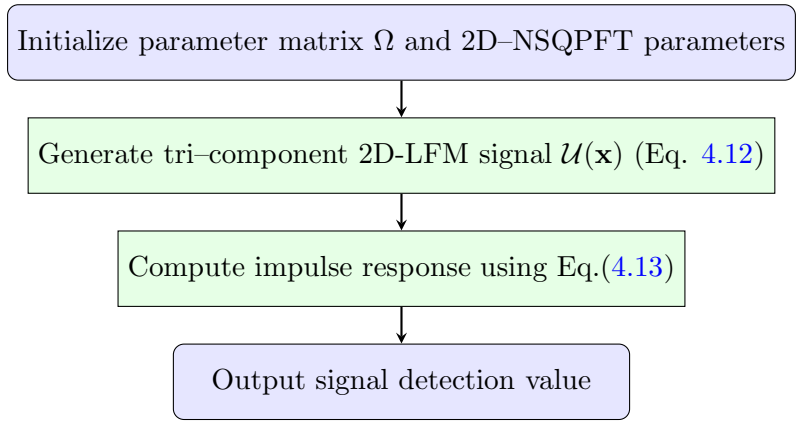
\begin{figure}[htbp]
\centering
\begin{tikzpicture}[node distance=1.5cm]

\node (step1) [startstop] { Initialize parameter matrix $\Omega$ and 2D--NSQPFT parameters};

\node (step2) [process, below of=step1] { Generate tri--component 2D-LFM signal $\mathcal{U}(\bf x)$ (Eq. $\ref{S4E12}$)};

\node (step3) [process, below of=step2] { Compute impulse response using Eq.$\eqref{S4E13}$};

\node (step4) [startstop, below of=step3] { Output signal detection value};

\draw [arrow] (step1) -- (step2);
\draw [arrow] (step2) -- (step3);
\draw [arrow] (step3) -- (step4);
\end{tikzpicture}
\caption{Flowchart showing the sequential computational stages for tri-component 2D-LFM signal detection using the 2D-NSQPWD.}
\label{fig:2DNSQPWD_flow}
\end{figure}
With the parameter settings $\kappa_n   \in \left\{5,\ 1,\ 11\right\}$, $\alpha_n   \in \{2,\ 4,\ 6\}$, $\beta_n   \in \left\{0.05,\ 0.05,\ 0.05\right\}$, $\mu_n \in \left\{0.15,\ 6,\ 0.25\right\}$, $\lambda_n \in \left\{0.04,\ 0.04,\ 0.04\right\}$ for $n\in \left\{1\dots 3\right\}$ and $T=40$, the transformation set $\Omega_1$ is defined as \begin{align*}
	\Omega_1 = \Bigg\{ & A = 
	\begin{bmatrix}
		1 & \frac{-1}{7} \\
		\frac{1}{7} & 1
	\end{bmatrix}, \,
	B = 
	\begin{bmatrix}
		2 & 1 \\
		1 & 4
	\end{bmatrix},  \,
	C = 
	\begin{bmatrix}
		1 & \frac{-19}{5} \\
		\frac{19}{5} & 1
	\end{bmatrix}, \,
	D = 
	\begin{bmatrix}
		4 & 5 \\
		0 & 7
	\end{bmatrix}, \,
	E = 
	\begin{bmatrix}
		2 & 7 \\
		2 & 5
	\end{bmatrix}
	\Bigg\}.
\end{align*}

\parindent=0mm\vspace{.1in}
 The 2D-NSQPWD of the tri-component signal $\mathcal{U}({\bf x})=\mathcal{U}_1({\bf x})+\mathcal{U}_2({\bf x})+\mathcal{U}_3({\bf x}) $ is depicted in
Fig.~\ref{F4}. The left column (Figs. 5(a), 5(c), 5(e)) shows the 3D time–frequency energy surfaces, revealing the joint time–frequency–energy structure of the signal, while the right column (Figs. 5(b), 5(d), 5(f)) presents the corresponding contour maps, offering a top-down view that emphasizes precise energy localization. Across all cases, the LFM components form clear diagonal ridges in the time–frequency plane, with slopes tied to chirp rates and shifts caused by changes in ${\bf x}$. Contour plots show sharp, high-energy bands aligned with these ridges, demonstrating the 2D-NSQPWD’s strong resolution and effective cross-term suppression. As ${\bf x}$increases, spectral support broadens slightly, affecting resolution. Overall, the 2D-NSQPWD reliably localizes energy and separates overlapping LFM components, even at moderate SNR levels.These observations confirm the method’s effectiveness in handling complex, noisy signals. The precise ridge delineation facilitates accurate parameter estimation for each component, while cross-term suppression reduces interference and improves clarity. This confirms the 2D-NSQPWD as a powerful and versatile tool for time-frequency analysis of multi-component LFM signals, effective even with closely spaced components where traditional methods struggle. Furthermore, the tunable parameter ${\bf x}$ allows balancing resolution and noise resilience for diverse applications.  
\begin{figure}[htbp]
	\centering		
	\begin{minipage}[b]{0.49\textwidth}
		\centering
		\includegraphics[width=\textwidth]{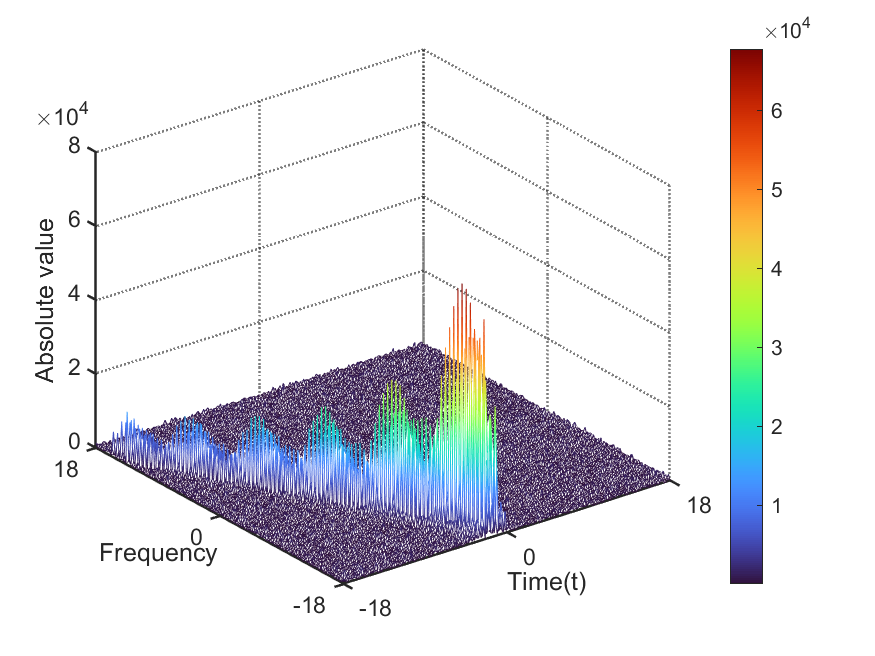}
		{\rm (a) $\mathcal{W}_\mathcal{U}^{\Omega_0}(6,5, \omega_1,\omega_2)$.}
	\end{minipage}
	\hfill
	\begin{minipage}[b]{0.49\textwidth}
		\centering
		\includegraphics[width=\textwidth]{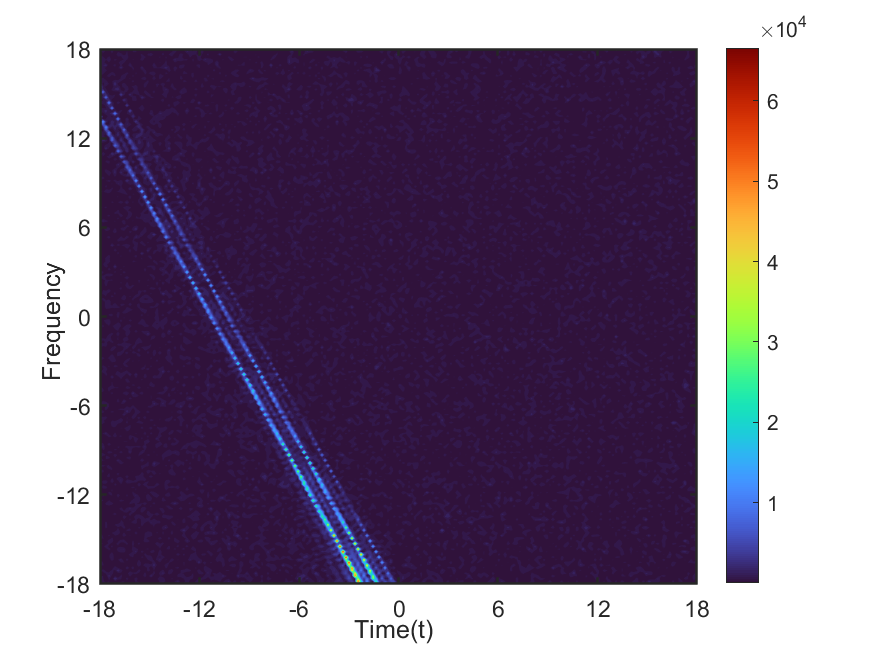}
		{\rm (b)  $\mathcal{W}_\mathcal{U}^{\Omega_0}(6,5, \omega_1,\omega_2)$.}
	\end{minipage}
\end{figure}
\begin{figure}[htbp]
	\centering		
	\begin{minipage}[b]{0.49\textwidth}
		\centering
		\includegraphics[width=\textwidth]{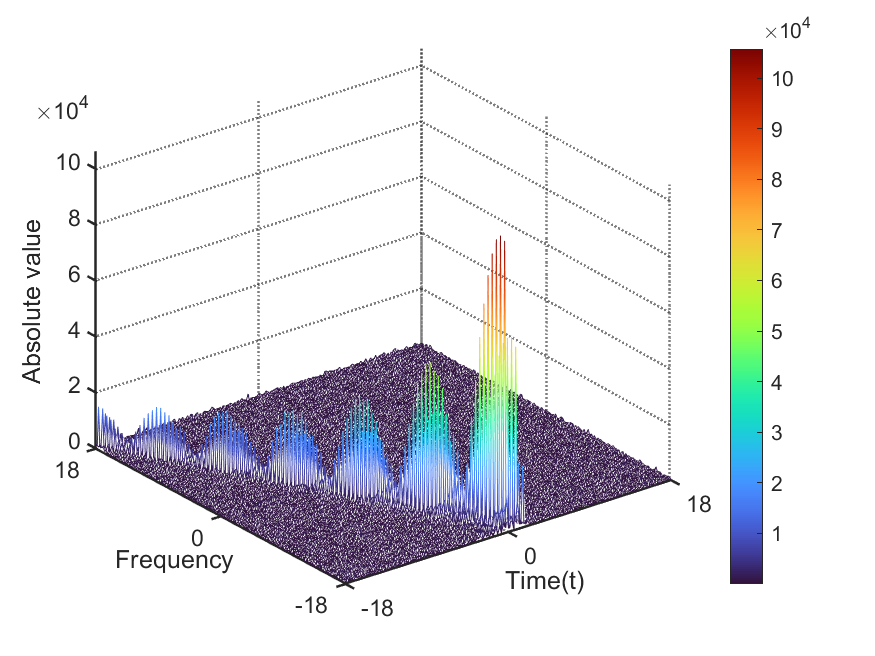}
		{\rm (c) $\mathcal{W}_\mathcal{U}^{\Omega_0}(4,5, \omega_1,\omega_2)$.}
	\end{minipage}
	\hfill
	\begin{minipage}[b]{0.49\textwidth}
		\centering
		\includegraphics[width=\textwidth]{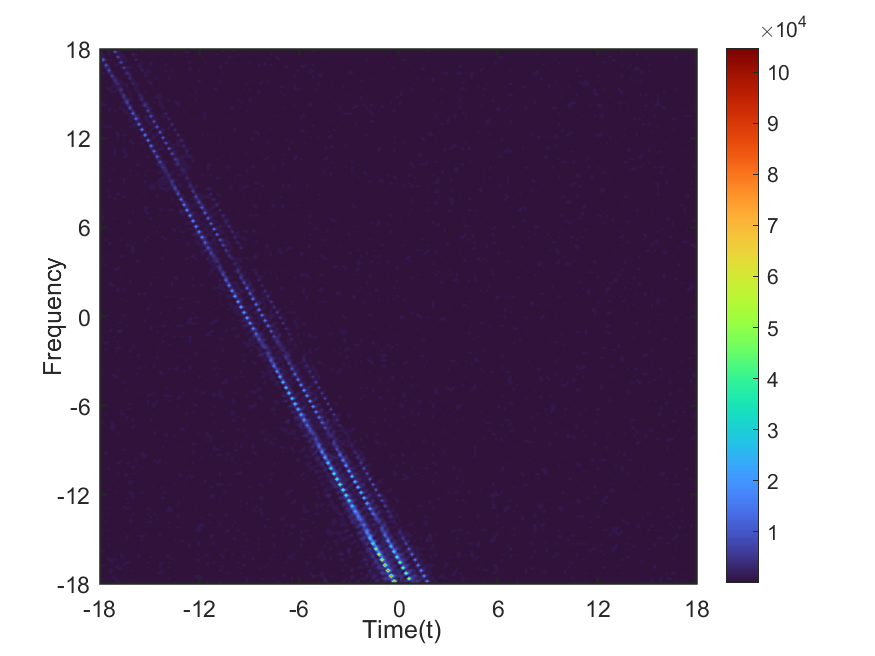}
		{\rm (d) $\mathcal{W}_\mathcal{U}^{\Omega_0}(4,5, \omega_1,\omega_2)$.}
	\end{minipage}
\end{figure}
\begin{figure}[htbp]
	\centering		
	\begin{minipage}[b]{0.49\textwidth}
		\centering
		\includegraphics[width=\textwidth]{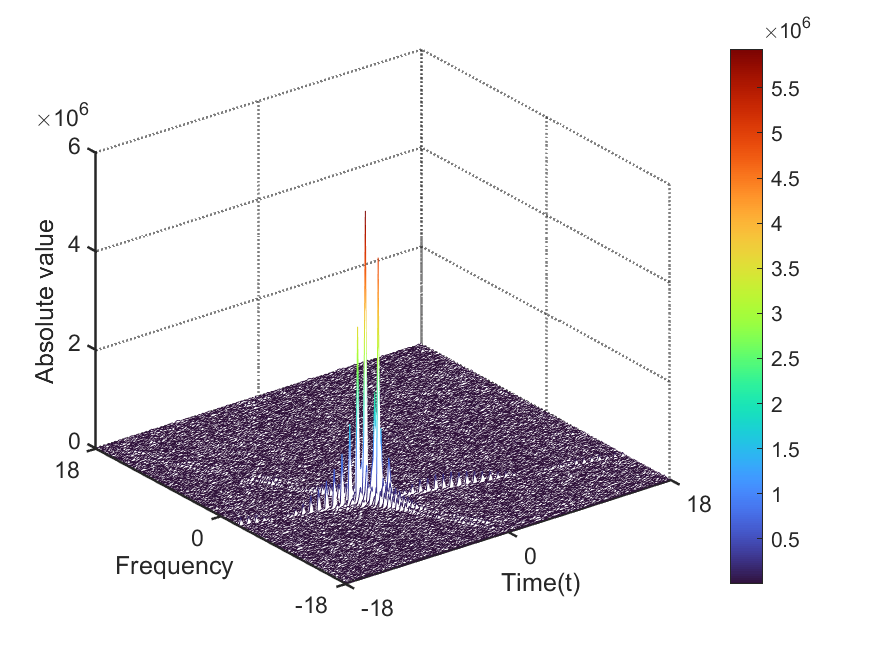}
		{\rm (e) $\mathcal{W}_\mathcal{U}^{\Omega_0}(4,1, \omega_1,\omega_2)$.}
	\end{minipage}
	\hfill
	\begin{minipage}[b]{0.49\textwidth}
		\centering
		\includegraphics[width=\textwidth]{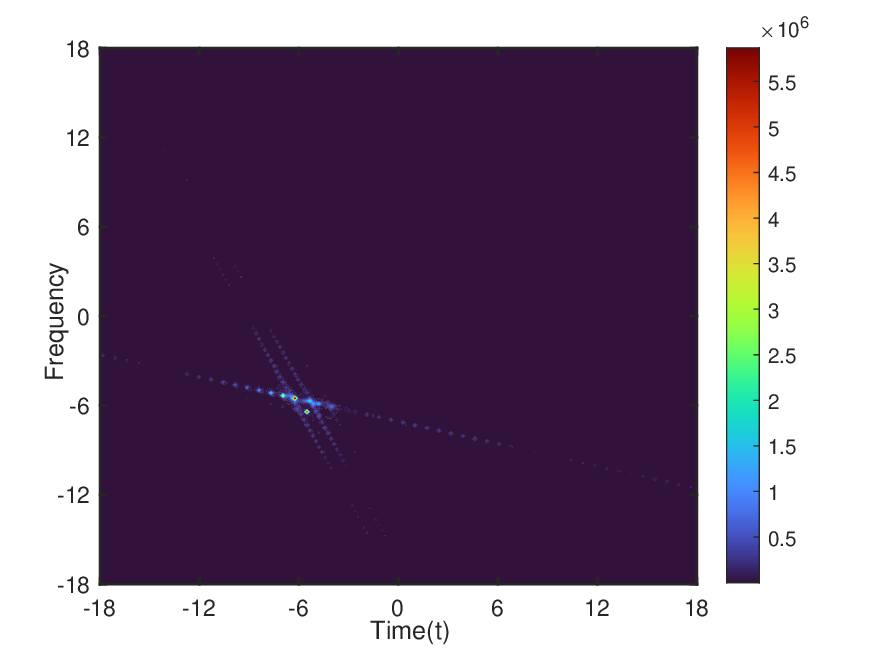}
		{\rm (f) $\mathcal{W}_\mathcal{U}^{\Omega_0}(4,1, \omega_1,\omega_2)$.}
	\end{minipage}
	\caption{\small The 2D-NSQPWD and corresponding contour plots of the tri-component signal $\mathcal{U}({\bf x})$ presented at an SNR of 10 dB for various values of $ \bf x $ and matrix parameters $\Omega_1$. }
	\label{F4}
\end{figure}

\section{Conclusion}\label{S5}

\parindent=0mm\vspace{.0in}
In this paper, we developed a novel  Wigner distributions linked to the 2D-NSQPFT framework and thoroughly established several fundamental properties of these new 2D integral transforms. These include shift invariance, conjugate symmetry, marginal distributions,Convolution, the Moyal identity, and their connections to the 2D-STFT and dilation behavior. As a key application, we demonstrated effective detection methods for single-, bi-, and tri-component 2D linear frequency modulated (LFM) signals using the proposed transforms, showcasing their practical utility in complex signal analysis.

\section*{Declarations}

\noindent\textbf{Data availability:} Data sharing is not applicable to this article as no datasets were generated or analyzed during the current study.

\vspace{1mm}
\noindent\textbf{Conflict of interest:} The authors declare that they have no competing interests.

\vspace{1mm}
\noindent\textbf{Author contributions:} All authors contributed equally to this work.

\vspace{1mm}
\noindent\textbf{Acknowledgments:} The first author is supported by University Grants Commission (UGC), under Ref No.221610002999  and second author is supported by Council of Scientific and Industrial Research (CSIR), Government of India, under file No. 09/1023(19437)/2024-EMR-I.


\begin{thebibliography}{0}
	
	
	\bibitem{Castro1}L.P. Castro,  M.R. Haque, M.M. Murshed, S. Saitoh, N.M. Tuan,  Quadratic Fourier transforms, Ann. Funct. Anal. 5, 10–23 (2014).
	 
	 \bibitem{Castro2}L.P. Castro, L.T. Minh, N.M. Tuan, New convolutions for quadratic-phase Fourier integral operators and their applications, Mediterr. J. Math. 15, 1–17 (2018).
	 	
	 \bibitem{Prasad}A. Prasad, P. B. Sharma, The quadratic-phase Fourier wavelet transform, Math. Methods Appl. Sci. 43, 1953–1969 (2020).
	 	
	 \bibitem{Shah1} F.A. Shah, W.Z. Lone, A.Y. Tantary, Short-time quadratic-phase Fourier transform, Optik. Int. J. Light Electron Opt. 245, 167689 (2021).
	 
	  \bibitem{Shah2}  W.Z. Lone, F.A. Shah, Shift-invariant spaces and dynamical sampling in quadratic-phase Fourier domains, Optik 260, 169063 (2022).
	  
	  \bibitem{AK} B. Gupta, A.K. Verma,  Short-time quaternion quadratic-phase Fourier transform and its uncertainty principles, Adv. Appl. Clifford Algebras 34, 28 (2024) \url{https://doi.org/10.1007/s00006-024-01334-x}. 
	  
	  \bibitem{CMA} L.P. Castro, R.C. Guerra, Multidimensional quadratic-phase Fourier transform and its uncertainty principles, Constr. Math. Anal. 8, 15-34 (2025).
	  
	  \bibitem{Bahri1}M. Bahri, E. Hitzer, R. Ashino,  R. Vaillancourt, Windowed Fourier transform of two-dimensional quaternionic signals, Appl. Math. Comput. 216, 2366–2379 (2010).
	  
	  \bibitem{Zayed1} A.I. Zayed, Two-dimensional fractional Fourier transform and some of its properties, Integral Transforms Spec. Funct. 29, 553–570 (2018).
	  
	  \bibitem{Pei1} S.C. Pei, J.J. Ding, Fractional Fourier transform: Wigner distribution and filter design for stationary and nonstationary random processes, IEEE Trans. Signal Process. 58, 4079–4092 (2001). 
	  
	  \bibitem{Dong}P. Dong, N.P. Galatsanos, Affine transformation resistant watermarking based on image
	  normalization, IEEE International Conference on Image Processing 3, 489–492 (2002).
	  
	  \bibitem{Zhao}L. Zhao, J.J. Healy, J.T. Sheridan, Two-dimensional nonseparable linear canonical transform, sampling theorem and unitary discretization, J. Opt. Soc. Am. A 31, 2631–2641 (2014).
	  
	  
	  \bibitem{Mukul} W.Z. Lone, M. Chauhan, A. K. Verma, A new class of short-time linear canonical transform: Theory and Applications, Math. Methods Appl. Sci. 1-17 (2025) \url{https://doi.org/10.1002/mma.11166}.
	   
	\bibitem{Wigner} E. Wigner, On the Quantum Correction For Thermodynamic Equilibrium, Phys. Rev. 40  749-759 (1932).
	
	\bibitem{Ville} J. Ville, Th\'eorie et applications de la notion de signal analytique, Cables et Transmissions. 2A,  61-74 (1948).
	
	\bibitem{Cohen} L. Cohen, Time-frequency distributions-a review, in Proceedings of the IEEE 7, 941-981 (1989).
		
	\bibitem{John} J.A. Johnston, Wigner distribution and FM radar signal design. IEE Proc. F Radar Signal Process. 136, 81-88 (1989).
		
	\bibitem{Bertrand} J. Bertrand and P. Bertrand, A class of affine Wigner functions with
		extended covariance properties, J. Math. Phys. 33, 2515–2527 (1992).
	
	\bibitem{Pei2} S.C. Pei, J.J. Ding, Two-dimensional affine generalized fractional Fourier transform, IEEE Transactions Signal Processing 49, 878–897 (2001).
	
		\bibitem{Sahin} A. Sahin, M.A. Kutay, H. M. Ozaktas, Non separable two-dimensional
	fractional Fourier transform, Appl. Opt. 23, 5444–5453,(1998).

	\bibitem{Zhang} Z. Zhang, Uncertainty Principle for Free Metaplectic Transformation, J Fourier Anal. Appl. 29, 71 (2023) \url{https://doi.org/10.1007/s00041-023-10052-0}. 
	
	\bibitem{Teali}  A.A. Teali,, F. A. Shah,  A.Y. Tantary,  Coupled fractional Wigner distribution with applications to LFM signals, Fractals 31, 2340020 (2023).
	
	
	\bibitem{Zayed2}A. Zayed,  A new perspective on the two- dimensional fractional Fourier transform and its relationship with the Wigner distribution, J. Fourier Anal. Appl. 25, 460–487 (2019).
	
	\bibitem{Minh1} L.T. Minh, Novel two-dimensional Wigner distribution and ambiguity function in the framework of the two-dimensional nonseparable linear canonical transform, Multidim. Syst. Sign. Process. 35, 11-35  (2024).
	
	\bibitem{Debnath1}L. Debnath, and B. Rao, On new two-dimensional Wigner-Ville nonlinear
	integral transforms and their basic properties, Integral Transform.
	Spec. Funct. 21, 165–174 (2010).
	
	\bibitem{Feng} Q. Feng, B.Z. Li, Convolution and correlation theorems for the two-dimensional linear canonical transform and its applications, IET Signal Process. 10, 125-132  (2016).

	
   \bibitem{Wei} D. Wei, Y. Shen, New two-dimensional Wigner distribution and ambiguity function associated with the two-dimensional nonseparable linear canonical transform, Circuits Syst. Signal Process. 41, 77-101 (2021).

 \bibitem{Gyrator} Z.J. Liu, H. Chen, T. Liu, P.F. Li, J.M. Dai, X.G. Sun, S.T. Liu, Double-image encryption based on the affine transform and the gyrator transform, J. Opt. 12, 035407(2010).

\bibitem{Lahiri}A. Lahiri, D. Kundu, A. Mitra, Efficient algorithm for estimating the parameters of two dimensional chirp signal, Sankhya B. 75, 65–89(2013).
    
\bibitem{Zhang1}Z. Zhang, Linear canonical Wigner distribution based noisy LFM signals detection through the output SNR improvement analysis, IEEE Trans. Signal Process. 21, 5527-5542(2019).
 
\end{thebibliography}
\end{document}